\pdfoutput=1
\documentclass[fleqn,usenatbib]{mnras}

\usepackage{newtxtext,newtxmath}
\usepackage[T1]{fontenc}
\usepackage{ae,aecompl}

\usepackage{longtable}

\usepackage{graphicx}	% Including figure files
\usepackage{amsmath}	% Advanced maths commands
\usepackage{makecell}   % Double line within a table cell
\usepackage{subcaption} % For sub-figures
\title[Substructure of Perseus]{The substructure of the Perseus star forming region: A survey with \textit{Gaia} DR2}

\author[T. Pavlidou et al.]{
Tatiana Pavlidou,$^{1}$\thanks{E-mail: tp32@st-andrews.ac.uk}
Aleks Scholz,$^{1}$
Paula S. Teixeira$^{1}$
\\
$^{1}$School of Physics and Astronomy, University of St Andrews\\
}

\date{Accepted XXX. Received YYY; in original form ZZZ}

\pubyear{2020}

\begin{document}
\label{firstpage}
\pagerange{\pageref{firstpage}--\pageref{lastpage}}
\maketitle

%papers to refer to!!:
% https://iopscience.iop.org/article/10.3847/1538-3881/abc0e6/pdf
% https://iopscience.iop.org/article/10.1088/0004-6256/150/3/95/pdf

\begin{abstract}
We use photometric and kinematic data from \textit{Gaia} DR2 to explore the structure of the star forming region associated with the molecular cloud of Perseus. Apart from the two well known clusters, IC\,348 and NGC\,1333, we present five new clustered groups of young stars, which contain between 30 and 300 members, named Autochthe, Alcaeus, Heleus, Electryon and Mestor. We demonstrate these are co-moving groups of young stars, based on how the candidate members are distributed in position, proper motion, parallax and colour-magnitude space. By comparing their colour-magnitude diagrams to isochrones we show that they have ages between 1 and 5\,Myr. Using 2MASS and WISE colours we find that the fraction of stars with discs in each group ranges from 10 to $\sim 50$ percent. The youngest of the new groups is also associated with a reservoir of cold dust, according to the Planck map at 353\,GHz. We compare the ages and proper motions of the five new groups to those of IC\,348 and NGC\,1333. Autochthe is clearly linked with NGC\,1333 and may have formed in the same star formation event. The seven groups separate roughly into two sets which share proper motion, parallax and age: Heleus, Electryon, Mestor as the older set, and NGC\,1333, Autochthe as the younger set. Alcaeus is kinematically related to the younger set, but at a more advanced age, while the properties of IC\,348 overlap with both sets. All older groups in this star forming region are located at higher galactic latitude.

\end{abstract}


%%%%%%%%%%%%%%%%%%%%%%%%%%%%%%%%%%%%%%%%%%%%%%%%%%

%%%%%%%%%%%%%%%%% BODY OF PAPER %%%%%%%%%%%%%%%%%%

\section{Introduction}
\label{intro}

%Young stellar clusters and their natal clouds constitute the best laboratories for research in star formation. Our solar system however is not in an ideal position for that: the nearest regions with ongoing star formation are at least 100\,pc away. Prior to the \textit{Gaia} mission, this was too distant to measure parallaxes for the overwhelming majority of the newly born stars in the solar vicinity. Similarly, prior to \textit{Gaia} the proper motions of young low-mass stars were in nearby star forming regions were often too small to be used for identifying co-moving objects. As a result, the census and the characterisation of the nearby star forming regions, especially outside the well known rich embedded clusters, remained incomplete. 

The Perseus star forming region is one of the largest associations of young stars within a distance of 500\,pc. This complex encompasses two well-studied young clusters, IC\,348 and NGC\,1333, and a population of distributed young stars, all embedded in the Perseus molecular cloud at distances of around 300\,pc \citep{bally_2008}. The three dimensional structure of the region is still poorly understood. Thanks to \textit{Gaia} DR2, the distances to IC348 and NGC\,1333 are now much better constrained to 320$\pm$26\,pc and 293$\pm$22\,pc \citep{ortiz_leon_2018}. 

Previous studies of the Perseus complex focus mostly on these two famous clusters, and in some cases, smaller clouds in between and around them. The canonical overview paper by \cite{bally_2008} discusses the two main clusters, the Barnard clouds B1 and B5 and the clouds L1448 and L1445, along with some individual sources, all located in an area approximately between and around the clusters (see their Figure 6). The Spitzer C2D survey, an extensive mid-infrared view of the region, covers predominantly the same regions \citep{jorgensen_2006}. In the submm, the James Clerk Maxwell Telescope (JCMT) survey by \cite{hatchell_2013} observed a restricted region around NGC\,1333, to construct dust temperature maps using Sub-millimetre Common-User Bolometer Array 2 (SCUBA-2) data. 

The most recent census of the stellar/substellar population was provided by \cite{luhman_2016}, hereafter referred to as L16. They compile a list of members for both IC\,348 and NGC\,1333 using optical and near-infrared spectra. Based on various indicators of youth, it has been surmised that IC\,348 is slightly older than NGC\,1333 with ages between 2-6\,Myr and 1\,Myr respectively (L16) in agreement across the literature (for example \cite{scholz_2012,stelzer_2012}). 

So far the areas beyond these inner parts of the complex have not been studied systematically at multiple wavelengths and with sufficient depth to identify young low-mass stars. In this paper we present the first results of a \textit{Gaia} DR2-based study of the entire Perseus star forming region. In particular, we report the discovery and characterisation of five new groups of young stars with ages 1-5\,Myr in this region and a few dozen to a few hundred potential members, which have not previously been reported in the literature.

We divide this paper as follows: In Section \ref{data} we introduce the data we use for our analysis, in Section \ref{selection} we select candidate members for the entire cloud of Perseus, in Section \ref{new_clusters} we present our selection method of the five new groups and the final lists of their candidate members. In Section \ref{youth} we demonstrate that these are groups of young stars and point out their relative age sequence. We discuss our results in Section \ref{discussion} and summarise in Section \ref{summary}.

\section{Data used in this paper}
\label{data}

Launched in 2013, \textit{Gaia} is considered the astrometric successor of \textit{Hipparcos} \citep{gaia_main_2016}. The second data release, \textit{Gaia} DR2, has enabled improvement in a large variety of astrophysical fields. The typical position, proper motion, and parallax uncertainties in DR2 are a few milli-arcsec for the faintest objects ($G> 21$ mag) and reduce to less than 0.1\,mas for bright stars \citep{gaia_dr2_2018}. The revolutionary precision in astrometry in combination with the depth allows us for the first time to study the three-dimensional distribution of the low-mass members of nearby star forming regions.

In order to identify objects with circumstellar discs we also use data from the Wide-field Infrared Survey Explorer (\textit{WISE}), observing at 4 bands centered at 3.4, 4.6, 12 and 22 $\mu$m and covering the entire sky \citep{wright_2010}. We use data from the AllWISE catalogue and combine it with Two Micron All Sky Survey \textit{2MASS} magnitudes \citep{skrutskie_2006} to search for excess emission in the infrared as an indicator of the presence of a disc.

Finally, we use the large-scale map of the dust emission measured at 353 GHz, as made available by the \textit{Planck} survey \citep{planck_2016}, to study the link between young stars and the presence of cold dust.

\section{Selecting young stars in Perseus}
\label{selection}

We use the \textit{Gaia} DR2 catalogue to identify young stellar objects (YSOs) in the Perseus star forming complex. We limit our survey to a $13\times 13$ degree area encompassing the two main clusters, IC\,348 and NGC\,1333. This survey area includes the entire previously known star forming complex \citep{bally_2008}.  

To define the selection criteria in parallax and proper motion, we use the confirmed members of the two known clusters, as listed in the comprehensive photometric and spectroscopic survey paper by L16. These known young stars, as observed by \textit{Gaia}, are constrained in proper motion to $+1 \ldots +10$ mas/yr in right ascension, $\alpha$, and $-15 \ldots -2$ mas/yr in declination, $\delta$. We use these ranges in proper motion as limits for our initial sample.

To confine our initial sample further, we also apply a cut-off in the parallax angle. The likely distance to the Perseus cloud ranges from 294 to 350\,pc according to \cite{zucker_2018}. We intentionally choose a wider range to explore the full depth of the star forming complex. \textit{Gaia} sources at these distances come with high uncertainties in parallax, therefore adopting a wider range ensures that our initial cloud sample is complete, and no real members are discarded. We keep sources in a range between 1.8 and 4.3 mas in parallax which translates to distances between $\sim$230 and 550\,pc. We also impose a limit on the parallax error of $<0.7$ mas which is the typical error at G = 20 mags (see Table 3 in \cite{gaia_dr2_2018}).

\begin{figure}
	\includegraphics[width=\columnwidth]{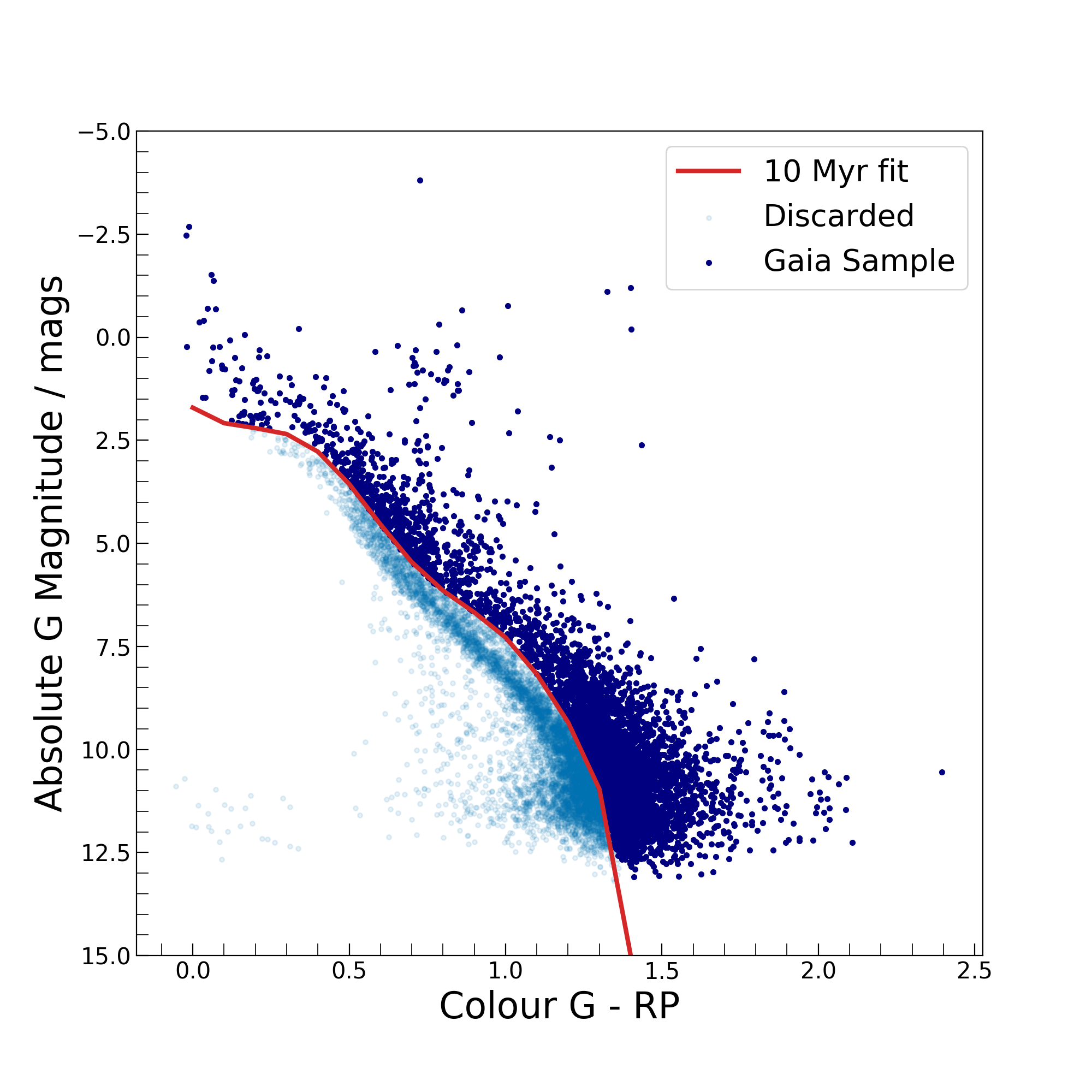}
    \caption{Colour-Magnitude Diagram from \textit{Gaia} photometry for our \textit{Gaia sample} (dark blue). Overplotted is a 10\,Myr isochrone constructed using models from \citet{marigo_2013}. The sources discarded are estimated to be older than 10\,Myr (light blue).}
    \label{fig:age_cut}
\end{figure}

Finally, we apply a conservative cut in the \textit{Gaia} (G-RP,G) colour-magnitude space (see Fig.~\ref{fig:age_cut}). To define this limit we use the 10\,Myr isochrone calculated by \citet{marigo_2013}. The cutoff implies that we keep all objects with an estimated age younger than 10\,Myr. The 10\,Myr isochrone is fitted here with a 5-degree polynomial. 
%using the Python numpy.polyfit function. 

The sample selected with these criteria in proper motion, parallax and age comprises in total 6202 objects and is hereafter called the {\it Gaia sample}. This is designed to be an inclusive sample, but it will contain contaminating stars in foreground and background of the star forming regions. Most of this population will require additional validation to be confirmed as YSOs and/or as members of Perseus. The full list of criteria used to select the {\it Gaia sample} is summarised in Table \ref{tab:perseus_conditions}.

\begin{figure}
	\includegraphics[width=\columnwidth]{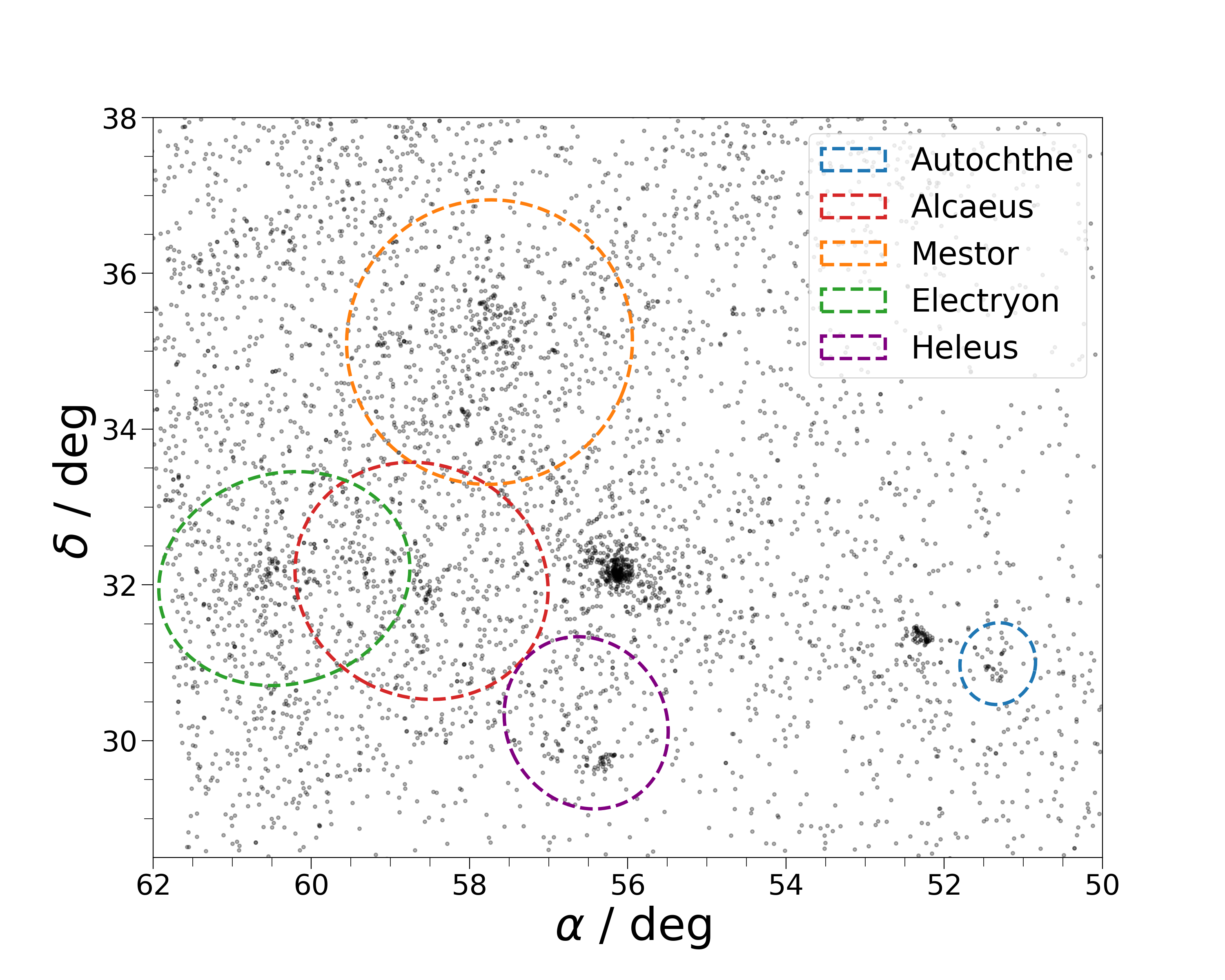}
    \caption{Spatial distribution of the 6202 sources in the \textit{Gaia sample}, satisfying the conditions of Table \ref{tab:perseus_conditions}. The ellipses show the 3$\sigma$ borders for each of the newly found groups.}
    \label{fig:gaia_sample}
\end{figure}

In Fig.~\ref{fig:gaia_sample} we show a map of the 6202 sources in the {\it Gaia sample}. The two known clusters NGC\,1333 and IC\,348 stand out as obvious dense populations around the central coordinates of ($52.3,31.3$) deg and ($56.1,32.2$) deg, respectively. Our basic search recovers 50\% and 78\% of the members listed in L16 for NGC\,1333 and IC348 respectively. 
The reason known members are not identified is that some of them are too faint to be detected by \textit{Gaia}.

\begin{table}
    \centering
	\caption{Perseus \textit{Gaia Sample} Selection Conditions}
	\begin{tabular}{l c} % four columns, alignment for each
    \hline
    \hline
    Property & Condition \\ [0.5ex] 
    \hline
    Area (sqdeg) & $ 13 \times 13 $ \\
    Proper Motion $\alpha$ (mas\,yr$^{-1}$) & $1.0<\mu_{\alpha}<10.0$ \\ 
    Proper motion $\delta$ (mas\,yr$^{-1}$)  & $-15.0<\mu_{\delta}<-2.0$ \\
    Parallax Angle (mas) & $  1.8<\varpi<4.3$ \\ 
    Approximate Age (Myr) & $\lesssim 10 $ \\ 
    Number of Sources & 6202  \\
    \hline
    \hline
	\end{tabular}
	\label{tab:perseus_conditions}
\end{table}

\begin{figure}
	\includegraphics[width=\columnwidth]{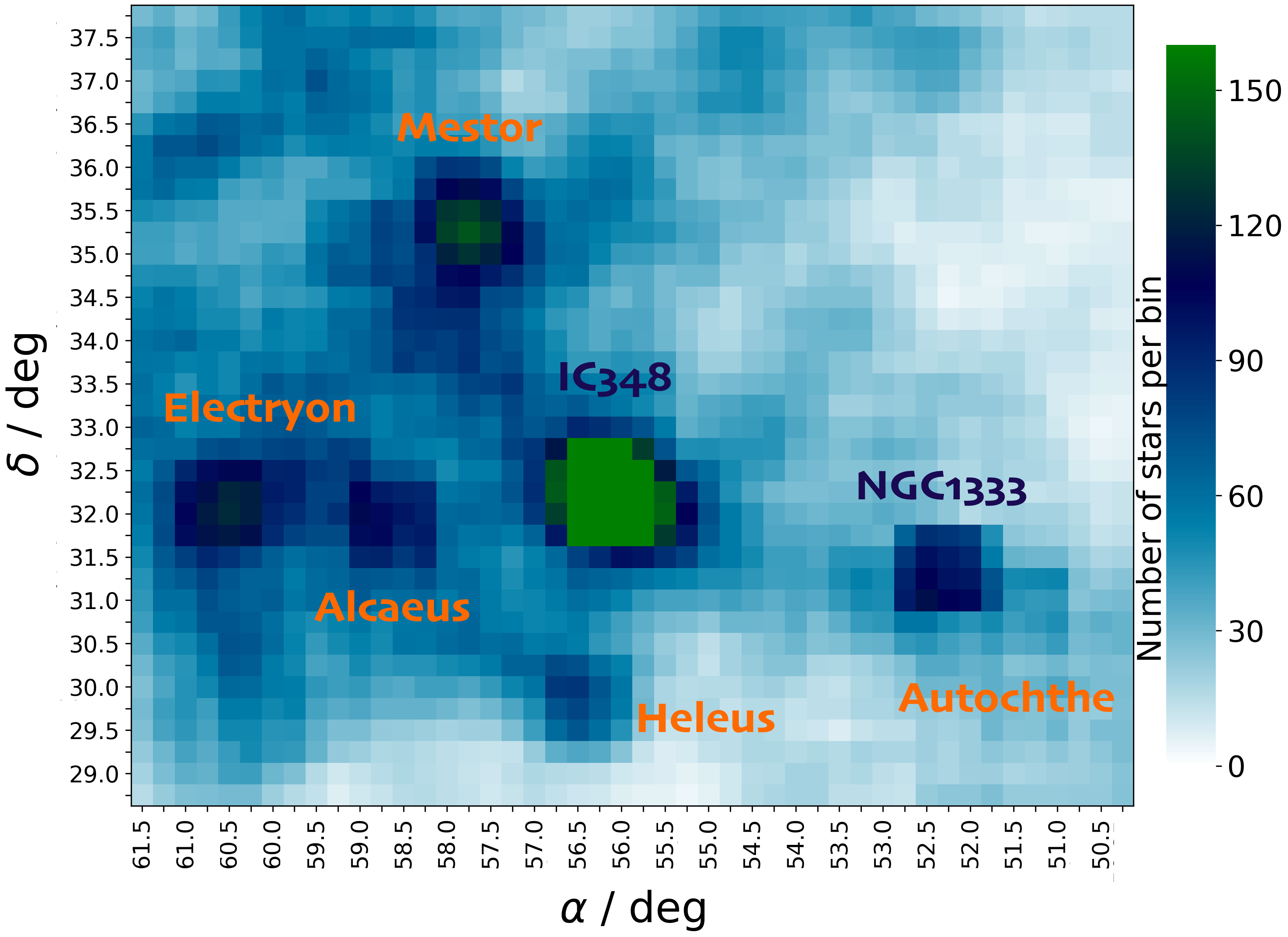}
    \caption{Number density plot of the \textit{Gaia sample}. To construct this plot, we use a box size used of $0.5 \times 0.5$ degrees and a stepsize of 0.25\,degrees.  The orange labels identify the five new groups.}
    \label{fig:density_map_per}
\end{figure}

\section{New groups of young stars}
\label{new_clusters}

\subsection{Identification of new groups}

Visual inspection of Fig.~\ref{fig:gaia_sample} shows several additional groupings of potential YSOs in the {\it Gaia sample}, outside the main clusters IC\,348 and NGC\,1333. We identify five groups, named Autochthe (PST1), Alcaeus (PST2), Mestor (PST3), Electryon (PST4) and Heleus (PST5) after five (out of the many) children of king Perseus in Greek Mythology. We also introduce the designation PST (last name initials of the authors) 1 to 5 for these five groups. We indicate the positions and names of these five groups in Fig.~\ref{fig:gaia_sample}.

To confirm the visual identification of these groups, we show a colour map of the number density of stars in the {\it Gaia sample} in Fig.~\ref{fig:density_map_per}. This figure was produced by counting the number of stars in each $0.5 \times 0.5$ degree box, while moving the centre of the box by a stepsize of 0.25\,degrees in $\alpha$ and $\delta$. Again, IC\,348 and NGC\,1333 clearly show up as regions of enhanced stellar density. In addition to IC\,348 and NGC\,1333, four of the five new groups of young stars mentioned above are distinctly visible as regions with a high density of stars and are labelled in the figure. Autochthe, located to the west of NGC\,1333, is not clearly distinguished from its neighbouring larger cluster.
We determine the peaks of the density enhancements in Fig.~\ref{fig:density_map_per} to define the spatial centers of the new groups. 
%The central coordinates for each group are listed in Table \ref{tab:cluster_conditions}. 
 
\subsection{Selection of members in spatial distribution}
\label{position_selection}

In the following, we define preliminary membership lists for the five new groups from the {\it Gaia sample}. Based on Fig.~\ref{fig:density_map_per} the groups have a diameter of around 1.0\,deg, with the exception of Autochthe, which is more compact. We define 5 square boxes, centered on the adopted central coordinates, with a sidelength chosen to encompass the visible density enhancement. Then we fit a Bivariate Gaussian to the $\alpha$-$\delta$ space of the objects contained in each box, using the tasks \texttt{fit\_bivariate\_normal} from the astroML.stats package in python and \texttt{\_bivariate\_normal} from astroML.stats.random \citep{VanderPlas_2012}. The Gaussian fit gives updated central coordinates and a measure for the dispersion of the members, $\sigma$. We keep as candidate members the sources that are located within a radius of 3$\sigma$ from the new central coordinates. We choose the 3$\sigma$ level as this corresponds well with the visual clustering in spatial distribution. This threshold presents a good compromise between completeness and avoiding excessive contamination. This step gives us preliminary lists of candidate members for each group. The circles with 3$\sigma$ radius are shown in Fig.~\ref{fig:gaia_sample}.

\begin{figure*}
    \includegraphics[width=\textwidth]{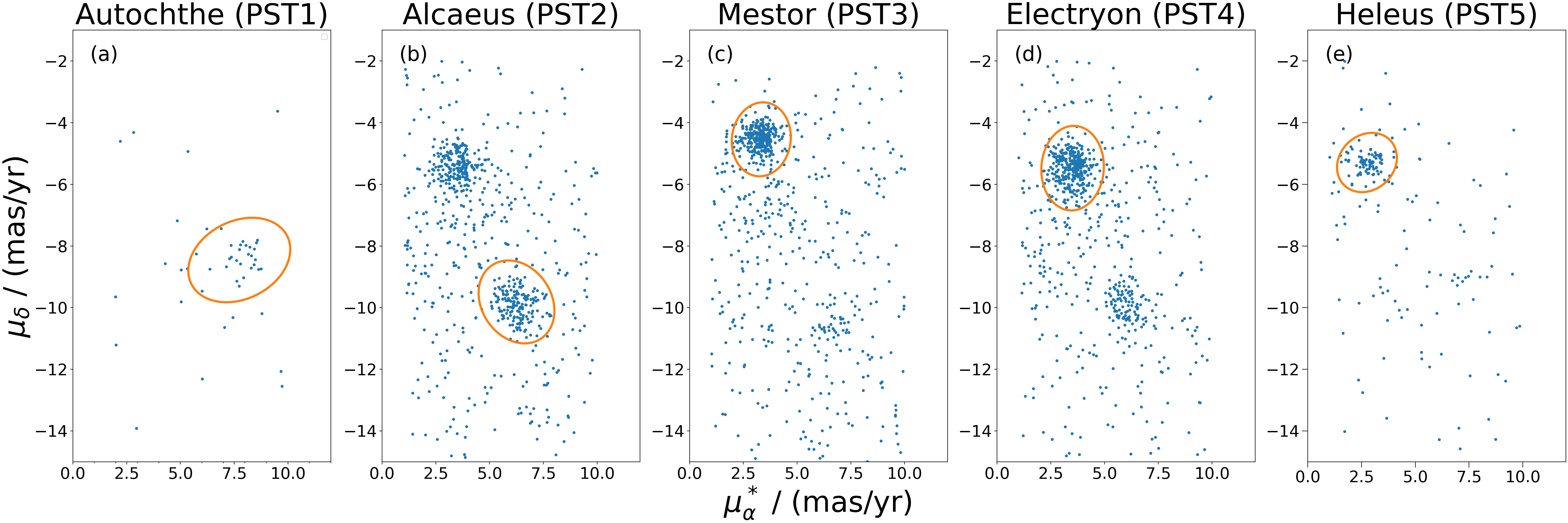}
    \caption{\textit{Gaia} proper motions of the initial samples for each of the 5 new groups. The orange ellipses show the 5$\sigma$ level within which we keep sources as members in each group. In panels (b) and (d) we see two clusterings, the bottom one corresponding to Alcaeus and the top one to Electryon. The two groups overlap in the plane of the sky but are well separated in proper motion and parallax.}
    \label{fig:pm_all}
\end{figure*}

\subsection{Selection of members using proper motion and parallax}
\label{pmotion_selection}

When plotted in proper motion space these previously defined samples are clearly clustered around a well-defined proper motion vector, with scattering much smaller than that of the {\it Gaia sample} (see Fig.~\ref{fig:pm_all}). This demonstrates that these five groups contain co-moving young stars within the wider Perseus star forming complex and hence constitute kinematically distinct groups.

To further confine the samples for each group, we define a box in proper motion space for each group and fit a Bivariate Gaussian, using the same method as for the spatial distribution. For all groups, we keep as members sources within 5$\sigma$ from the mean proper motion, shown as ellipses in Fig.~\ref{fig:pm_all}. This further confinement gives a new sample for each group. 

\begin{figure*}
	\includegraphics[width=\textwidth]{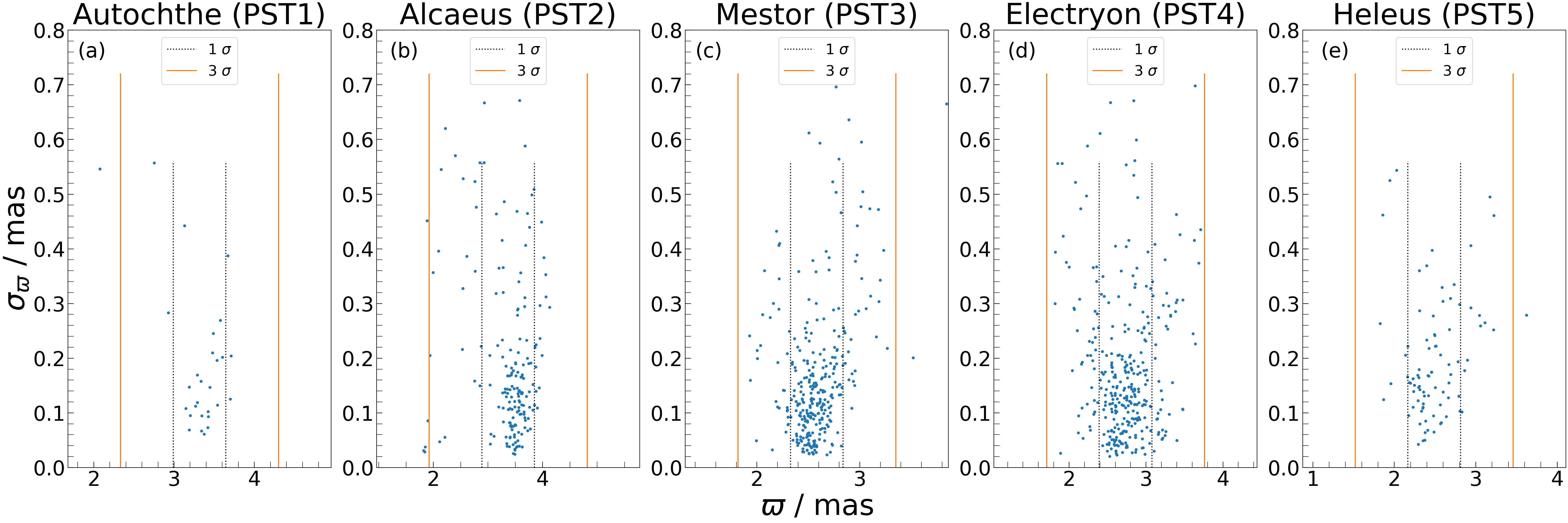}
    \caption{Parallax errors versus the parallax of the 5 new groups. Autochthe (PST1) and Alcaeus (PST2) are at parallax of $\sim3.3$ mas corresponding to a distance of around 300\,pc, similar to NGC\,1333. Mestor (PST3), Electryon (PST4) and Heleus (PST5) are all located further away at mean parallaxes of $\sim2.6$ mas ($\sim400$\,pc).}
    \label{fig:plx_panels}
\end{figure*}

Next we probe the distribution of parallaxes for each group. In Fig.~\ref{fig:plx_panels} we show the parallax errors against the parallax angles. Recall that all objects in the {\textit Gaia sample} satisfy only a very wide parallax selection criterion (see Table \ref{tab:perseus_conditions}). From Fig.~\ref{fig:plx_panels} it is evident that the members of each of the five new groups are centered around a well defined mean, i.e. they share a common distance. This provides a further argument in favour of these five groups being actual clusterings of young stars, as opposed to accidental density enhancements. Fig.~\ref{fig:plx_panels} also shows that candidate members of each group are mostly contained within 3$\sigma$ from the mean. We discard all sources outside 3$\sigma$ of the mean.  

Discarding the few parallax outliers gives us the final number of members for each group. The final samples sizes are 170 for Alcaeus, 302 for Mestor, 329 for Electryon, 85 for Heleus, and 27 for Autochthe. We cross-checked the list for Autochthe with the list of members for NGC\,1333 published by L16, and did not find any sources in common. All objects should be considered candidates prior to spectroscopic confirmation, but given that the final sample for each group shows consistent parallax and proper motion, as well as clustering on the sky, the contamination from foreground and background objects should be low. The selection criteria and sample sizes for all five groups are listed in Table \ref{tab:cluster_conditions}. In Appendix \ref{appendixB}, Tables \ref{tab:autochthe_members}-\ref{tab:heleus_members} list the members of the five groups with their key \textit{Gaia} parameters and their \textit{AllWISE} designation.

\begin{table*}
\centering
\caption{Selection criteria and sample properties for the five new groups.}
\label{tab:cluster_conditions}
\begin{tabular}{lccccc} % four columns, alignment for each
\hline
\hline
                                           & Autochthe (PST1) & Alcaeus (PST2) & Mestor (PST3) & Electryon (PST4) & Heleus (PST5)\\
\hline
($\alpha$,$\delta$)$_\mathrm{mean}$ (deg)  & 51.34, 30.95     & 58.53, 31.98  & 57.91, 35.09  & 60.35, 32.19     & 56.42, 29.83 \\ 
$\sigma(\alpha$,$\delta)$ (deg)            & 0.40             & 1.25           & 1.21          & 1.19             & 0.71  \\
\textit{Initial sample size}               & 47               & 718            & 704           & 739              & 184 \\

\hline
($\mu_{\alpha}^*$,$\mu_{\delta}$)$_\mathrm{mean}$ (mas\,yr$^{-1}$) & 7.7, -8.5 & 6.3, -9.8 & 3.3, -4.5 & 3.5, -5.5 & 2.8, -5.3 \\
$\sigma$($\mu_{\alpha}^*$,$\mu_{\delta})$ (mas\,yr$^{-1}$)         & 1.1       & 0.89      & 0.73      & 0.80      & 0.68 \\  
\textit{Sample size after proper motion cut}                       & 28        & 175       & 305       & 329       & 86 \\

\hline
$\varpi_\mathrm{mean}$ (mas)                & 3.3        & 3.4        & 2.6        & 2.7        & 2.5   \\  
$\sigma(\varpi)$ (mas)                      & 0.33       & 0.48       & 0.26       & 0.34       & 0.32 \\
Mean distance (pc)                          & 298$\pm24$ & 291$\pm13$ & 395$\pm16$ & 370$\pm15$ & 413$\pm31$ \\
\textit{Final sample size}                  & 27         & 170        & 302        & 329        & 85   \\ 

\hline
Number of WISE counterparts                 & 24   & 161  & 291  & 311  & 82 \\
Percentage of sources with discs            & 56\% & 12\% & 23\% & 18\% & 24\%\\
\hline
\hline

\end{tabular}
\end{table*}

\section{Evidence for Youth}
\label{youth}

In this section, we aim to demonstrate that these are indeed groups of young stars, which will lead to an estimate of the ages. 

\subsection{Colour-Magnitude Diagrams}
\label{cmds}

\begin{figure*}
	\includegraphics[width=\textwidth]{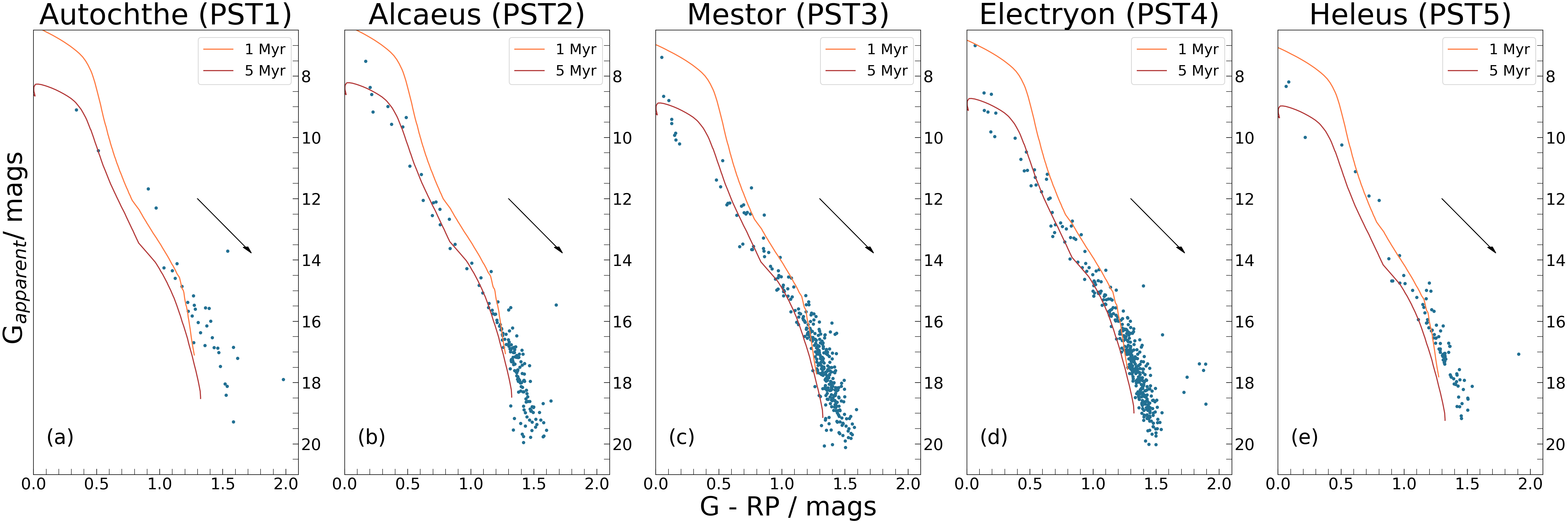}
    \caption{Colour-Magnitude Diagrams from \textit{Gaia} photometry for our final samples for each group. The red and orange solid lines correspond to isochrones of 1\,Myr and 5\,Myr respectively adopted from \citet{marigo_2013}. The isochrones were shifted to each group's distance as listed in Table \ref{tab:cluster_conditions}. The black arrow corresponds to an extinction of A$_\mathrm{v}$ = 2.0 mag, calculated for G and $G_{RP}$ \textit{Gaia} bands using information from \citet{wang_2019}.}
    \label{fig:cmds}
\end{figure*}

In Fig.~\ref{fig:cmds} we present the colour-magnitude diagrams in \textit{Gaia} photometry of the final samples of the five groups including 1\,Myr and 5\,Myr isochrones as calculated from \cite{marigo_2013}, using a solar metallicity, Z $\sim$ 0.0152. We shift the isochrones to the mean distance of each group as given in Table \ref{tab:cluster_conditions}. We note that the distance error will somewhat affect the position of the isochrones; for a distance of $\sim$300\,pc and an error of $\sim$20\,pc the isochrone position along the y-axis is uncertain by $\sim$ 0.3\,mags. We include a reddening vector in the diagrams.

All groups form a clear sequence in colour-magnitude space in line with the expected positions of young stars, confirming their youth and confirming they are coeval groups. Comparing with the isochrones, most of Autochthe members and a few of Heleus members seem to be very young, with ages of 1\,Myr or less. Alcaeus, Mestor and Electryon show a wider range in age up to 5\,Myr. In particular, most members of Alcaeus are located around the 5\,Myr isochrone, making Alcaeus the oldest of the five groups. 

The masses for the isochrones range from $\sim0.1$ to a few $M_{\odot}$. As it can be seen in Fig.~\ref{fig:cmds}, our samples do contain some objects with masses below the low mass limit of the isochrones. The fact that isochrones are close to the sequences of young stars indicates low extinction and a small range in extinctions. The only exception is Autochthe which may be affected by up to 2\,mag in optical extinction. 

\begin{figure}
	\includegraphics[width=\columnwidth]{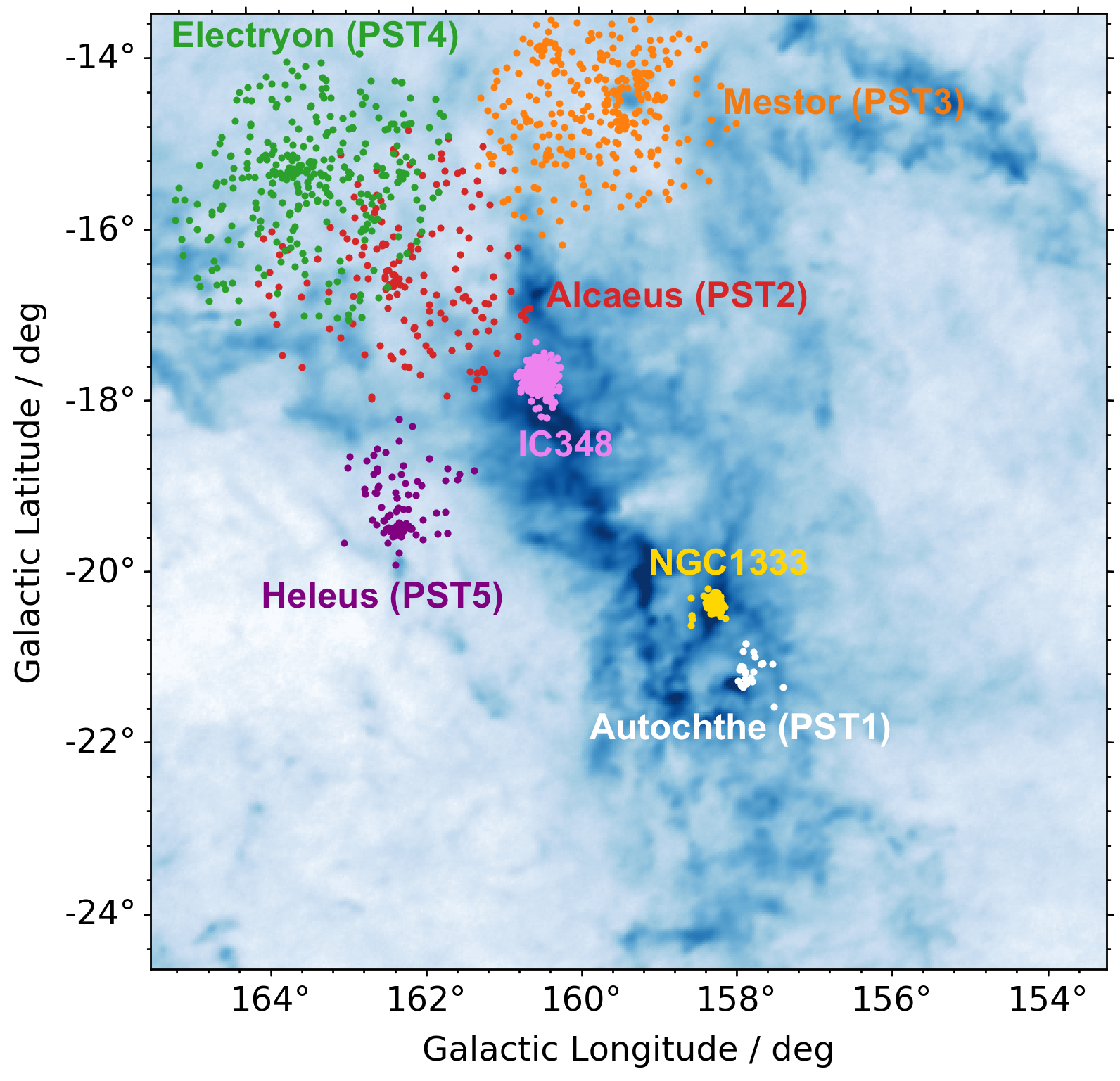}
    \caption{A dust emission map at 353\,Ghz from Planck data of the whole Perseus star forming complex \citep{planck_2016}. The samples of young stars in the five groups discussed in this paper, plus the two known clusters, are overplotted.}
    \label{fig:planck}
\end{figure}

\subsection{Infrared Excess Emission}
\label{IR_excess}

To test the evolutionary state of the candidate members we also search for infrared excess. We use infrared data from the WISE mission since none of the five groups are covered by Spitzer \citep{jorgensen_2008}. We cross-match our samples for each group with the AllWISE catalogue using a position tolerance of 1.0\,arcsec. The overwhelming majority of the sources have a WISE counterpart. We list the number of counterpart sources found for each group in Table \ref{tab:cluster_conditions}. 

\begin{figure*}
	\includegraphics[width=\textwidth]{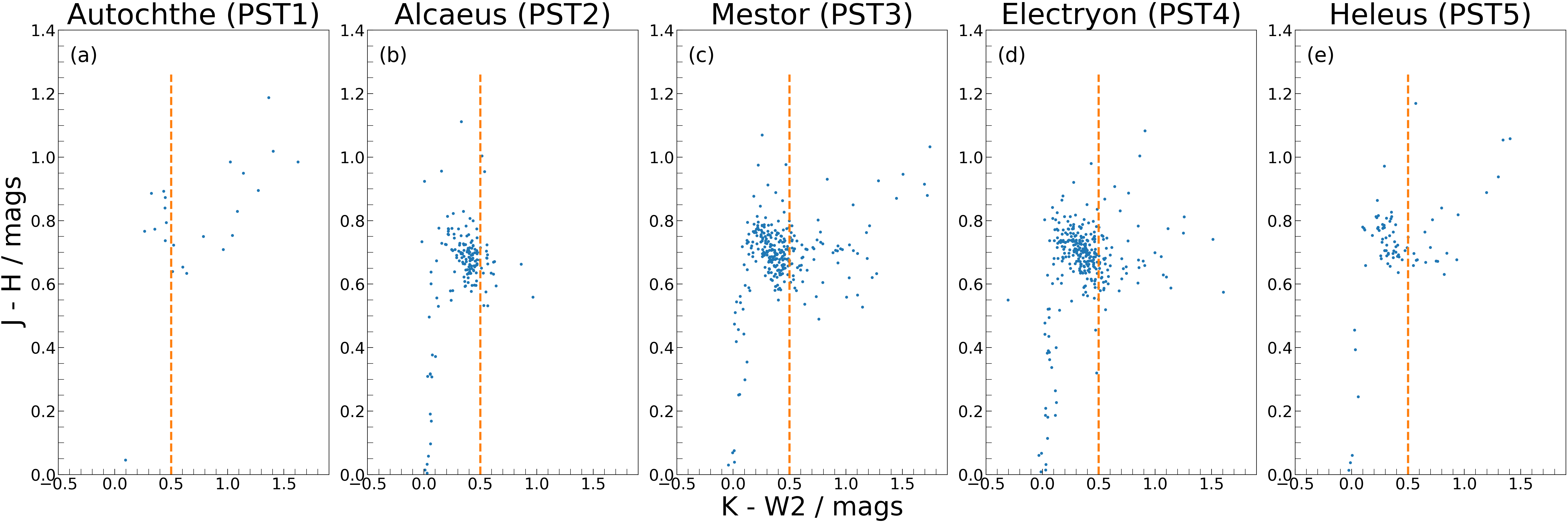}
    \caption{Infrared colour-colour plot for all five groups. Objects to the right of the vertical line at $K - W2 = 0.5$ mag are sources with infrared excess, likely due to the presence of a disc \citep[e.g.][]{teixeira_2020}.}
    \label{fig:all_wise}
\end{figure*}

In Fig.~\ref{fig:all_wise} we show a colour-colour plot for each group using 2MASS and WISE data. In this figure objects with excess emission due to circumstellar dusty discs will appear on the right hand side, with a colour of $K-W2>0.5$. We adopt this value from \cite{teixeira_2012}, Figure 7 (top panel), where stars with thick discs all have colours $K-[4.5 \mu m]$ larger than $\sim$0.5. According to this criterion, a subset of the members in our groups shows evidence for the presence of a disc. Autochthe hosts the largest percentage of objects with discs, with 15 out of its 27 members ($\sim$56\%). Heleus and Mestor have similar disc fractions, 20 out of 85 ($\sim$24\%) and 69 out of 302 ($\sim$23\%), respectively. For Electryon, 58 out of 329 ($\sim$18\%) stars show infrared excess as defined above. Alcaeus has the lowest disk fraction with 20 out of 170 ($\sim$12\%). Using the prevalence of discs as indicator for age, Autochthe is the youngest of the groups, and Alcaeus the oldest, in line with the estimate from the colour-magnitude diagrams.

\subsection{Location within the molecular cloud}
\label{cloud_location}

In Fig.~\ref{fig:planck} we show the dust emission map of the Perseus molecular cloud at 353GHz available from the Planck survey, along with the samples of the five new groups in different colours. The two known clusters, IC\,348 and NGC\,1333, are located on the broad cloud emission band. Autochthe appears as a very compact clustering and is clearly located in a region of dust emission as well. The remaining groups sit outside the main band. However, Mestor's central region is coinciding with less pronounced dust emission. Heleus is comparatively close to the large dust clouds in the complex. Alcaeus and Electryon do not coincide with dust emission. We also note that Electryon, Alcaeus, and Mestor are large dispersed groups, while Heleus and Autochthe are more compact, similar to the known clusters. Overall, the links with the dust emission and the physical structure of the new groups broadly confirm the evolutionary sequence suggested by the analysis in Sect. \ref{cmds} and \ref{IR_excess}, with Autochthe as youngest, Alcaeus and Electryon as oldest groups. We note that the fact that we see groups of stars in the same position as a dust clump does not necessarily mean that the young stars are physically associated with the clump, since we do not know the distance to the clumps.

\begin{figure}
	\includegraphics[width=\columnwidth]{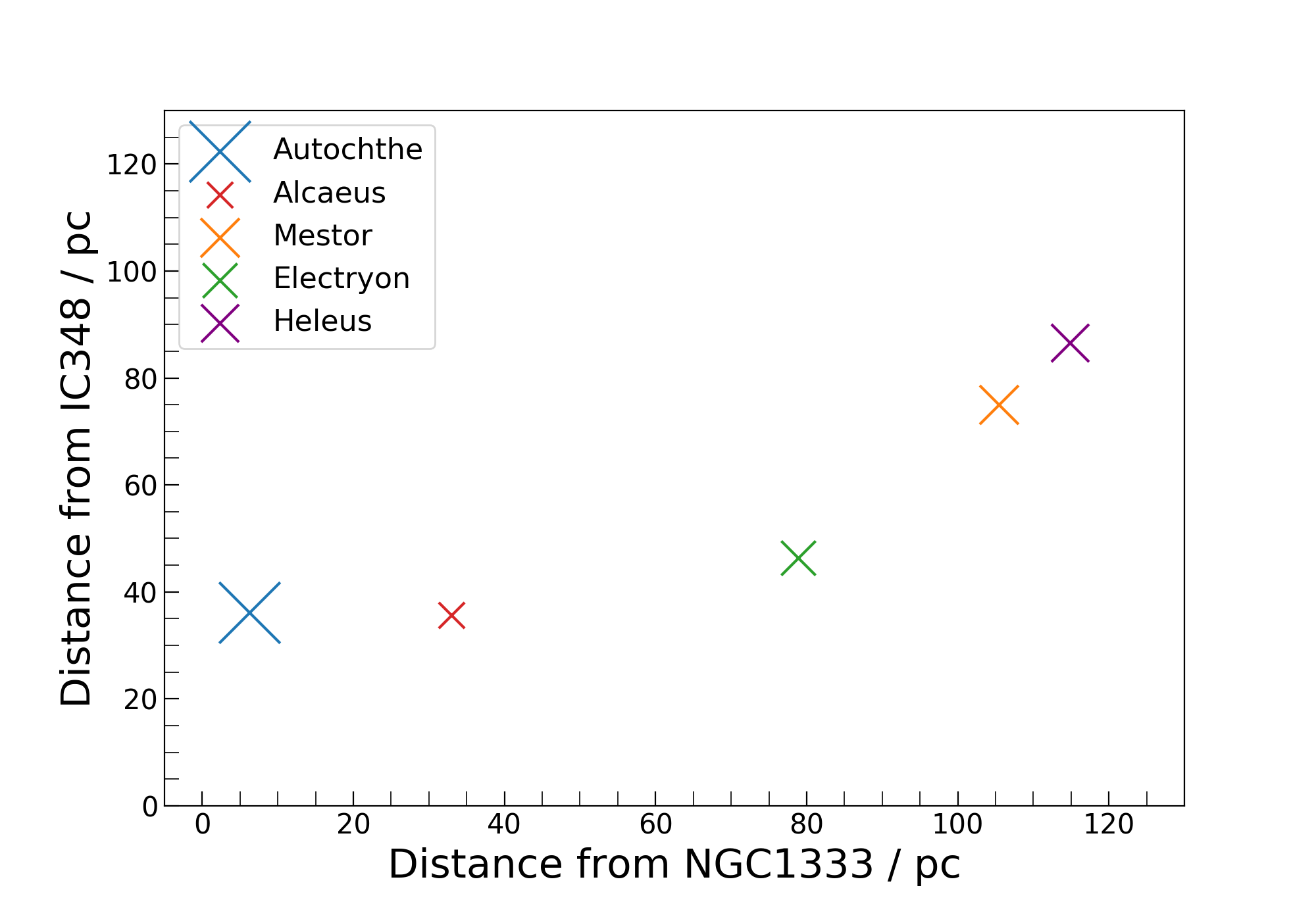}
    \caption{Three-dimensional distance of each of the five new groups from IC\,348 versus their distance form NGC\,1333. The size of the symbol is proportional to the percentage of sources with discs.}
    \label{fig:rel_dist}
\end{figure}

\begin{figure}
	\includegraphics[width=\columnwidth]{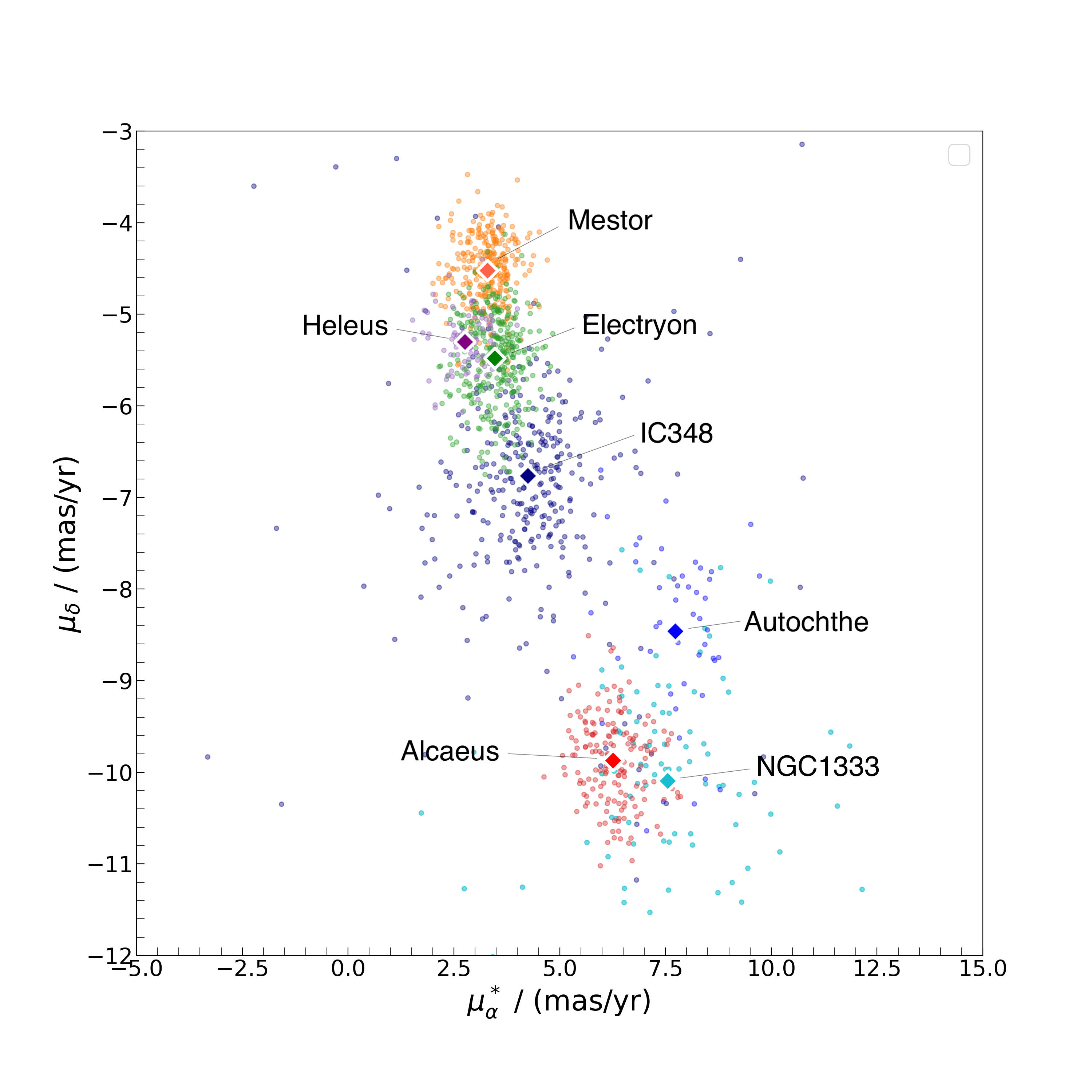}
    \caption{The proper motions of all seven groups associated with Perseus. The IC\,348 and NGC\,1333 samples were cross-matched between the membership list of L16 and the \textit{Gaia} DR2 catalogue. We show the mean proper motion vector of each of the groups with diamonds.}
    \label{fig:7_pms}
\end{figure}

\begin{figure*}
	\includegraphics[width=\textwidth]{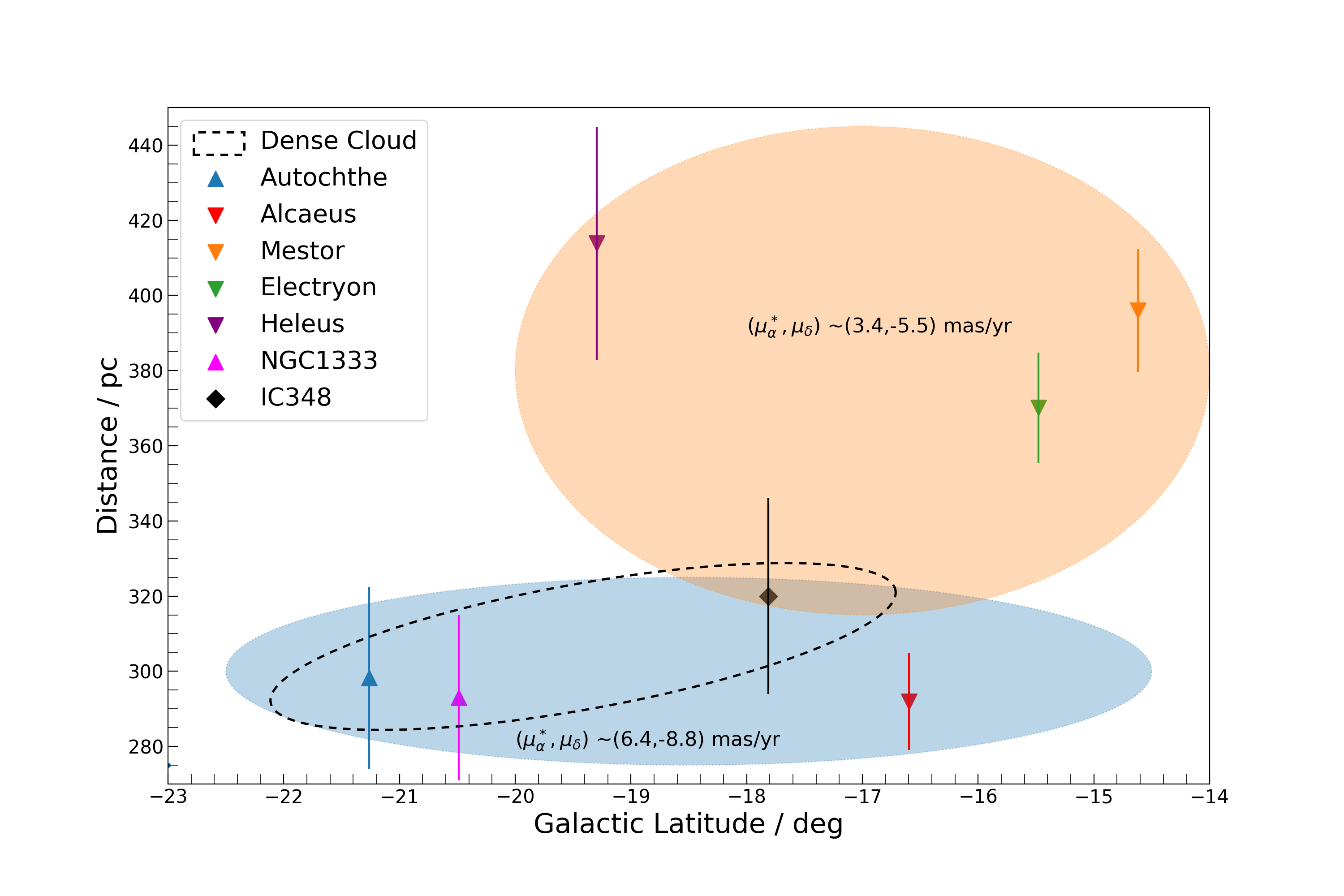}
    \caption{Distance versus Galactic Latitude for the seven groups associated with the Perseus star-forming complex. 
    The upward pointing triangle symbols represent groups with an age of about 1\,Myr, the downward pointing triangle symbols represent groups older than 3\,Myr, whilst the diamond symbol represents IC\,348 with an age spread of 2 to 6\,Myr (L16). The seven groups separate into two kinematic sets according to their proper motion values (the mean proper motion of each set is written in the corresponding ellipse). We also include a schematic representation of the dense molecular cloud based on \ref{fig:planck}.}
    \label{fig:cluster_groups}
\end{figure*}

\section{Discussion: links between centres of star formation}
\label{discussion}

The existence of five clustered groups of young stars in the Perseus complex, in addition to the known clusters NGC\,1333 and IC\,348, enables new insights into the star formation history of the region. In this section we explore the links between all seven groups. We base the discussion on the position, proper motion, distance and age of each group. We introduce membership samples for the two known clusters, by cross-matching the membership lists from L16, to our knowledge the best available census for these clusters, with the {\it Gaia} DR2 catalogue to end up with 101 and 375 members for NGC\,1333 and IC\,348, respectively. 

We note that, we refer to Autochthe, Alcaeus, Mestor, Electryon and Heleus as ``groups" and to NGC\,1333 and IC\,348 as ``clusters" throughout the paper. Across the literature (e.g. L16) NGC\,1333 and IC\,348 are referred to as clusters. Most of the five new groups we present are more spread out in spatial distribution than the clusters, with the exception of Autochthe which harbours significantly fewer members.

\subsection{Trends in location}
\label{spatial}

In the $\alpha$-$\delta$ plane, Mestor, Heleus, Electryon, Alcaeus and IC\,348 are located on the eastern side of the cloud while Autochthe and NGC\,1333 are located further to the west. With respect to distances, NGC\,1333, Autochthe and Alcaeus sit at around 300\,pc, corresponding to the nearest portion of the cloud. Mestor, Heleus and Electryon are further away at distances between $\sim$ 370\,pc and $\sim$ 410\,pc.  IC\,348 is located in-between these two sets, at $\sim$ 320\,pc \citep{ortiz_leon_2018}. 

In Fig.~\ref{fig:rel_dist} we show the relative three-dimensional distances of all five groups from IC\,348 against their distances from NGC\,1333. The size of the crosses corresponds to the percentage content of stars with discs (see Sect. \ref{IR_excess}). Mestor, Heleus and Electryon are close to each other, but far away from the two known centres of star formation -- more than 80\,pc from NGC\,1333 and more than 40\,pc from IC\,348. All three also show low fractional content of stars with discs. On the other hand, Autochthe is the closest to both of the known clusters -- only $\sim 5$\,pc away from NGC\,1333 and $\sim 30$\,pc from IC\,348 -- and has the highest fraction of stars with discs. Finally, Alcaeus is equidistant from both NGC\,1333 and IC\,348 at $\sim 30$\,pc. Thus, in terms of spatial positioning, Mestor, Heleus and Electryon form one subset, and Autochthe, Alcaeus, NGC\,1333 and IC\,348 another. 

\subsection{Trends in kinematics}
\label{kinematics}

In Fig.~\ref{fig:7_pms} we show all seven groups in proper motion space, together with their mean proper motion as diamonds. In terms of proper motion, the groups again clearly separate into two sets. Mestor, Heleus, and Electryon are occupying the same proper motion space. NGC\,1333, Autochthe and Alcaeus are distinct in this diagram, but overlap with each other. In general, IC\,348 almost spans the whole proper motion space, and overlaps with Autochthe, but also with Electryon and Heleus.

\subsection{Ages}
\label{age}

We estimate the age sequence within the new five groups based on four different indicators (see Sect. \ref{youth}): trend in colour-magnitude diagrams, fraction of stars with discs, the groups' association with cold dust, and how compact the groups are. Taking into account these indicators together, there is a clear age sequence. 

Autochthe is clearly the youngest of the five new groups with a 56\% content of stars with discs and a strong association with dust (see Fig.~\ref{fig:planck}). Its brighter members also closely follow the 1\,Myr isochrone in the colour-magnitude diagram. That means, Autochthe is likely coeval to NGC\,1333, also at $\sim$ 1\,Myr according to L16. 

Mestor and Heleus also contain very young stars, according to the indicators listed above. About a fifth to a quarter of their members have discs. Some of their members appear close to or above the 1\,Myr isochrone in the colour-magnitude diagram. Their members form relatively compact groups as well, as seen from Fig.~\ref{fig:planck}. Electryon and Alcaeus are clearly older with the lowest disc fractions (18\% and 12\% respectively) and many members with magnitude and colours consistent with the 5\,Myr isochrone. They show no association with dust and are more spread out in spatial distribution.

Thus, the age sequence from youngest to oldest is: Autochthe and NGC\,1333, Heleus and Mestor, Electryon and Alcaeus. IC\,348 is assumed to host stars with a very wide range of ages, from 2 to 6\,Myr (L16), encompassing much of the age range of the new groups. 

\subsection{Star formation history}

With the parameters analysed above in sections \ref{spatial}, \ref{kinematics} and \ref{age}, a few clear links between separate groups of young stars become apparent. The most obvious one is between Autochthe and NGC\,1333: These two show similarities across all parameters, in particular, they are located very close together and may be part of the same star formation event. 

Apart from that, Mestor, Electryon and Heleus show similar proper motions, similar range of ages, and are all in the north-eastern part of the complex in terms of their spatial location. All three are off the main cloud. Alcaeus is a special case, kinematically and spatially close to NGC\,1333, but with an age range similar to Electryon. IC\,348 spans a wide range of proper motions and ages which almost covers that of all the rest of the groups. It is the richest cluster and clearly the most pronounced centre of star formation in the complex.

Only the pair NGC\,1333 and Autochthe, as well as IC\,348 show evidence for ongoing star formation. Thus, in the Perseus cloud the centres of current star formation are located at the near side of the cloud, at distance around 300\,pc, both on the eastern (with IC\,348) and western side (with NGC\,1333 and Autochthe). The more distant groups are all more evolved and have stopped actively forming stars. All groups, with the exception of Autochthe, show an age spread of at least 2\,Myr in the colour-magnitude diagrams, pointing to extended periods of star formation bursts throughout the region. 

In Fig.~\ref{fig:cluster_groups} we aim to visualise the set of properties for the seven groups of young stars in Perseus. The groups and clusters are plotted in distance vs. Galactic Latitude. The figure shows again that two sets of groups share a similar distance. The segregation in proper motions between the two sets is illustrated with the two ellipses of different colours. The three new groups in the orange ellipse, Electryon, Mestor and Heleus, also share a similar age -- triangles pointing down in that diagram represent ages of 3-5\,Myr. In the blue ellipse, Autochthe and NGC\,1333 located close to each other again share a similar and very young age of $\sim$1\,Myr (shown in triangles pointing up). Alcaeus, which is also in this ellipse due to its proper motion and distance, is an older group ($\sim$5\,Myr). IC\,348 lies in between these two sets and it is plotted as a black diamond symbol to represent the range of ages it spans. The current star forming cloud is indicated in a dashed-line ellipse containing NGC\,1333, Autochthe and IC\,348. 

Fig.~\ref{fig:cluster_groups} illustrates that the older groups in Perseus are located closer to the Galactic Plane, at lower latitudes. Mestor, Electryon, and Alcaeus at latitudes of $-14$ to $-17$ degrees have formed first, about 5\,Myr ago, so has Heleus at $-19$ degrees. Star formation is finished in these regions. NGC\,1333 and Autochthe, at higher latitudes of $-20$ to $-22$ degrees are sites of active ongoing star formation, with stars at typical ages of 1\,Myr.

\section{Summary}
\label{summary}

We summarize below the main conclusions of our study:
\begin{enumerate}
    \item We report the discovery of five new groups in the Perseus star-forming complex: Autochthe, Alcaeus, Mestor, Electryon and Heleus.
    \item Four of the new groups are located off-cloud: Heleus, Electryon, Mestor (at distances between 380\,pc and 420\,pc) and Alcaeus (at a distance of 290\,pc);
    \item All the off-cloud groups have ages between 3 and 5\,Myr, and disc fractions between 12 and 24\%. The on-cloud group, Autochthe, has an age of $\lesssim$1\,Myr and a disc fraction of 56\%;
    \item Autochthe is located to the east of NGC\,1333, at a distance of 298\,pc, similar to NGC\,1333 at 293$\pm$22\,pc. The two share very similar proper motions too, and are both around 1\,Myr old. These similarities in spatial distribution, proper motion and age suggest that these two may have formed together.  
    \item The newly discovered groups appear to segregate into two kinematic sets of similar proper motion; the first set is composed of Heleus, Electryon and Mestor with a mean proper motion of $(\mu_\alpha^*, \mu_\delta)$ = (3.5, -5.5) mas\,yr$^{-1}$, and the second set is composed of Alcaeus, Autochthe, and NGC\,1333 with a mean proper motion of $(\mu_\alpha^*, \mu_\delta)$ = (6.4, -8.8) mas\,yr$^{-1}$;
    \item We find a general sequence of star formation, where the older groups are located closer to the Galactic Plane, whereas the youngest groups are at higher Galactic latitudes.
    %\item We propose a hypothetical star formation scenario of the Perseus complex where each of these proper motion groups trace the original kinematics of two parental clouds, respectively. IC\,348 is located at the intersection of these clouds, which would account for its on-going star formation for the past 5\,Myr and a proper motion value that is in-between that of the two kinematic groups.
\end{enumerate}

\section*{Acknowledgements}

We thank the referee for a constructive report that helped to improve the paper. This project was supported by STFC grant ST/R000824/1.
This work has made use of data from the European Space Agency (ESA) mission
{\it Gaia} (\url{https://www.cosmos.esa.int/gaia}), processed by the {\it Gaia}
Data Processing and Analysis Consortium (DPAC,
\url{https://www.cosmos.esa.int/web/gaia/dpac/consortium}). Funding for the DPAC
has been provided by national institutions, in particular the institutions
participating in the {\it Gaia} Multilateral Agreement.
This research has made use of Python, https://www.python.org, NumPy \citep{vanderwalt11}, and Matplotlib \citep{hunter07}.
This research made use of APLpy, an open-source plotting package for Python and hosted at  http://aplpy.github.com \citep{robitaille12}. This research made use of Astropy, a community-developed core Python package for Astronomy \citep{robitaille13}. This  research  made  use  of  TOPCAT  (Tool for OPerations on Catalogues And Tables), an  interactive  graphical viewer  and  editor  for  tabular  data  \citep{taylor05}.

\bigskip

\textit{Note added in Proof:} Independently, \cite{kounkel_2019} have published a subsample of the young stars selected in this paper, as part of population numbered 21 and 77 in their catalogue (see also \cite{kounkel_2020}). These two populations cover a mix of members from various groups identified in our paper. Their population 21 contains 8 out of the 27 members in our sample of Autochthe and 86 out the 170 members in our sample for Alcaeus. Very few members of our sample of Electryon (3) and of our sample of Heleus (2) are also part of this population. Their population 77 contains 186 out of the 302 members in our sample of Mestor, 170 out of the 329 members in our sample of Electryon and 37 out of the 85 members in our sample of Heleus. 

\section*{Data Availability}
The data underlying this article are available in the article and in its online supplementary material. The datasets were derived from the public domain of Gaia, at \url{https://gea.esac.esa.int/archive/}, WISE, at \url{https://irsa.ipac.caltech.edu/Missions/wise.html} and Planck at \url{https://irsa.ipac.caltech.edu/data/Planck/release_2/all-sky-maps/}.

%\citep{daSilva_2006}

%%%%%%%%%%%%%%%%%%%%%%%%%%%%%%%%%%%%%%%%%%%%%%%%%%



\bibliographystyle{mnras}
%\bibliography{references} 

\begin{thebibliography}{}
\makeatletter
\relax
\def\mn@urlcharsother{\let\do\@makeother \do\$\do\&\do\#\do\^\do\_\do\%\do\~}
\def\mn@doi{\begingroup\mn@urlcharsother \@ifnextchar [ {\mn@doi@}
  {\mn@doi@[]}}
\def\mn@doi@[#1]#2{\def\@tempa{#1}\ifx\@tempa\@empty \href
  {http://dx.doi.org/#2} {doi:#2}\else \href {http://dx.doi.org/#2} {#1}\fi
  \endgroup}
\def\mn@eprint#1#2{\mn@eprint@#1:#2::\@nil}
\def\mn@eprint@arXiv#1{\href {http://arxiv.org/abs/#1} {{\tt arXiv:#1}}}
\def\mn@eprint@dblp#1{\href {http://dblp.uni-trier.de/rec/bibtex/#1.xml}
  {dblp:#1}}
\def\mn@eprint@#1:#2:#3:#4\@nil{\def\@tempa {#1}\def\@tempb {#2}\def\@tempc
  {#3}\ifx \@tempc \@empty \let \@tempc \@tempb \let \@tempb \@tempa \fi \ifx
  \@tempb \@empty \def\@tempb {arXiv}\fi \@ifundefined
  {mn@eprint@\@tempb}{\@tempb:\@tempc}{\expandafter \expandafter \csname
  mn@eprint@\@tempb\endcsname \expandafter{\@tempc}}}

\bibitem[\protect\citeauthoryear{{{Astropy Collaboration}} et~al.,}{{{Astropy
  Collaboration}} et~al.}{2013}]{robitaille13}
{{Astropy Collaboration}} et~al., 2013, \mn@doi [Astronomy and Astrophysics]
  {10.1051/0004-6361/201322068}, 558, A33

\bibitem[\protect\citeauthoryear{{Bally}, {Walawender}, {Johnstone}, {Kirk}  \&
  {Goodman}}{{Bally} et~al.}{2008}]{bally_2008}
{Bally} J.,  {Walawender} J.,  {Johnstone} D.,  {Kirk} H.,   {Goodman} A.,
  2008, {The Perseus Cloud}.
p.~308

\bibitem[\protect\citeauthoryear{{Gaia Collaboration} et~al.,}{{Gaia
  Collaboration} et~al.}{2016}]{gaia_main_2016}
{Gaia Collaboration} et~al., 2016, \mn@doi [\aap]
  {10.1051/0004-6361/201629272}, \href
  {https://ui.adsabs.harvard.edu/abs/2016A&A...595A...1G} {595, A1}

\bibitem[\protect\citeauthoryear{{Gaia Collaboration} et~al.,}{{Gaia
  Collaboration} et~al.}{2018}]{gaia_dr2_2018}
{Gaia Collaboration} et~al., 2018, \mn@doi [\aap]
  {10.1051/0004-6361/201833051}, \href
  {https://ui.adsabs.harvard.edu/abs/2018A&A...616A...1G} {616, A1}

\bibitem[\protect\citeauthoryear{{Hatchell} et~al.,}{{Hatchell}
  et~al.}{2013}]{hatchell_2013}
{Hatchell} J.,  et~al., 2013, \mn@doi [\mnras] {10.1093/mnrasl/sls015}, \href
  {https://ui.adsabs.harvard.edu/abs/2013MNRAS.429L..10H} {429, L10}

\bibitem[\protect\citeauthoryear{Hunter}{Hunter}{2007}]{hunter07}
Hunter J.~D.,  2007, \mn@doi [Computing in Science and Engineering]
  {10.1109/MCSE.2007.55}, 9, 90

\bibitem[\protect\citeauthoryear{{J{\o}rgensen} et~al.,}{{J{\o}rgensen}
  et~al.}{2006}]{jorgensen_2006}
{J{\o}rgensen} J.~K.,  et~al., 2006, \mn@doi [\apj] {10.1086/504373}, \href
  {https://ui.adsabs.harvard.edu/abs/2006ApJ...645.1246J} {645, 1246}

\bibitem[\protect\citeauthoryear{{J{\o}rgensen}, {Johnstone}, {Kirk}, {Myers},
  {Allen}  \& {Shirley}}{{J{\o}rgensen} et~al.}{2008}]{jorgensen_2008}
{J{\o}rgensen} J.~K.,  {Johnstone} D.,  {Kirk} H.,  {Myers} P.~C.,  {Allen}
  L.~E.,   {Shirley} Y.~L.,  2008, \mn@doi [\apj] {10.1086/589956}, \href
  {https://ui.adsabs.harvard.edu/abs/2008ApJ...683..822J} {683, 822}

\bibitem[\protect\citeauthoryear{{Kounkel} \& {Covey}}{{Kounkel} \&
  {Covey}}{2019}]{kounkel_2019}
{Kounkel} M.,  {Covey} K.,  2019, \mn@doi [\aj] {10.3847/1538-3881/ab339a},
  \href {https://ui.adsabs.harvard.edu/abs/2019AJ....158..122K} {158, 122}

\bibitem[\protect\citeauthoryear{{Kounkel}, {Covey}  \& {Stassun}}{{Kounkel}
  et~al.}{2020}]{kounkel_2020}
{Kounkel} M.,  {Covey} K.,   {Stassun} K.~G.,  2020, \mn@doi [\aj]
  {10.3847/1538-3881/abc0e6}, \href
  {https://ui.adsabs.harvard.edu/abs/2020AJ....160..279K} {160, 279}

\bibitem[\protect\citeauthoryear{{Lindegren} et~al.,}{{Lindegren}
  et~al.}{2018}]{lindegren_2018}
{Lindegren} L.,  et~al., 2018, \mn@doi [\aap] {10.1051/0004-6361/201832727},
  \href {https://ui.adsabs.harvard.edu/abs/2018A&A...616A...2L} {616, A2}

\bibitem[\protect\citeauthoryear{{Luhman}, {Esplin}  \& {Loutrel}}{{Luhman}
  et~al.}{2016}]{luhman_2016}
{Luhman} K.~L.,  {Esplin} T.~L.,   {Loutrel} N.~P.,  2016, \mn@doi
  [Astrophysical Journal] {10.3847/0004-637X/827/1/52}, \href
  {http://adsabs.harvard.edu/abs/2016ApJ...827...52L} {827, 52}

\bibitem[\protect\citeauthoryear{{Marigo}, {Bressan}, {Nanni}, {Girardi}  \&
  {Pumo}}{{Marigo} et~al.}{2013}]{marigo_2013}
{Marigo} P.,  {Bressan} A.,  {Nanni} A.,  {Girardi} L.,   {Pumo} M.~L.,  2013,
  \mn@doi [\mnras] {10.1093/mnras/stt1034}, \href
  {https://ui.adsabs.harvard.edu/abs/2013MNRAS.434..488M} {434, 488}

\bibitem[\protect\citeauthoryear{{Ortiz-Le{\'o}n} et~al.,}{{Ortiz-Le{\'o}n}
  et~al.}{2018}]{ortiz_leon_2018}
{Ortiz-Le{\'o}n} G.~N.,  et~al., 2018, \mn@doi [\apj]
  {10.3847/1538-4357/aada49}, \href
  {https://ui.adsabs.harvard.edu/abs/2018ApJ...865...73O} {865, 73}

\bibitem[\protect\citeauthoryear{{Planck Collaboration} et~al.,}{{Planck
  Collaboration} et~al.}{2016}]{planck_2016}
{Planck Collaboration} et~al., 2016, \mn@doi [\aap]
  {10.1051/0004-6361/201527101}, \href
  {https://ui.adsabs.harvard.edu/abs/2016A&A...594A...1P} {594, A1}

\bibitem[\protect\citeauthoryear{Robitaille \& Bressert}{Robitaille \&
  Bressert}{2012}]{robitaille12}
Robitaille T.,  Bressert E.,  2012, Astrophysics Source Code Library, p.
  ascl:1208.017

\bibitem[\protect\citeauthoryear{{Scholz}, {Muzic}, {Geers}, {Bonavita},
  {Jayawardhana}  \& {Tamura}}{{Scholz} et~al.}{2012}]{scholz_2012}
{Scholz} A.,  {Muzic} K.,  {Geers} V.,  {Bonavita} M.,  {Jayawardhana} R.,
  {Tamura} M.,  2012, \mn@doi [The Astrophysical Journal]
  {10.1088/0004-637X/744/1/6}, \href
  {http://adsabs.harvard.edu/abs/2012ApJ...744....6S} {744, 6}

\bibitem[\protect\citeauthoryear{{Skrutskie} et~al.,}{{Skrutskie}
  et~al.}{2006}]{skrutskie_2006}
{Skrutskie} M.~F.,  et~al., 2006, \mn@doi [\aj] {10.1086/498708}, \href
  {https://ui.adsabs.harvard.edu/abs/2006AJ....131.1163S} {131, 1163}

\bibitem[\protect\citeauthoryear{{Stelzer}, {Preibisch}, {Alexander},
  {Mucciarelli}, {Flaccomio}, {Micela}  \& {Sciortino}}{{Stelzer}
  et~al.}{2012}]{stelzer_2012}
{Stelzer} B.,  {Preibisch} T.,  {Alexander} F.,  {Mucciarelli} P.,  {Flaccomio}
  E.,  {Micela} G.,   {Sciortino} S.,  2012, \mn@doi [\aap]
  {10.1051/0004-6361/201118118}, \href
  {https://ui.adsabs.harvard.edu/abs/2012A&A...537A.135S} {537, A135}

\bibitem[\protect\citeauthoryear{Taylor}{Taylor}{2005}]{taylor05}
Taylor M.~B.,  2005, Astronomical Data Analysis Software and Systems XIV ASP
  Conference Series, 347, 29

\bibitem[\protect\citeauthoryear{{Teixeira}, {Lada}, {Marengo}  \&
  {Lada}}{{Teixeira} et~al.}{2012}]{teixeira_2012}
{Teixeira} P.~S.,  {Lada} C.~J.,  {Marengo} M.,   {Lada} E.~A.,  2012, \mn@doi
  [\aap] {10.1051/0004-6361/201015326}, \href
  {https://ui.adsabs.harvard.edu/abs/2012A&A...540A..83T} {540, A83}

\bibitem[\protect\citeauthoryear{{Teixeira}, {Scholz}  \& {Alves}}{{Teixeira}
  et~al.}{2020}]{teixeira_2020}
{Teixeira} P.~S.,  {Scholz} A.,   {Alves} J.,  2020, \mn@doi [\aap]
  {10.1051/0004-6361/201936756}, \href
  {https://ui.adsabs.harvard.edu/abs/2020A&A...642A..86T} {642, A86}

\bibitem[\protect\citeauthoryear{{VanderPlas}, {Connolly}, {Ivezic}  \&
  {Gray}}{{VanderPlas} et~al.}{2012}]{VanderPlas_2012}
{VanderPlas} J.,  {Connolly} A.~J.,  {Ivezic} Z.,   {Gray} A.,  2012, in
  Proceedings of Conference on Intelligent Data Understanding (CIDU. pp 47--54
  (\mn@eprint {arXiv} {1411.5039}), \mn@doi{10.1109/CIDU.2012.6382200}

\bibitem[\protect\citeauthoryear{{Wang} \& {Chen}}{{Wang} \&
  {Chen}}{2019}]{wang_2019}
{Wang} S.,  {Chen} X.,  2019, \mn@doi [\apj] {10.3847/1538-4357/ab1c61}, \href
  {https://ui.adsabs.harvard.edu/abs/2019ApJ...877..116W} {877, 116}

\bibitem[\protect\citeauthoryear{{Wright} et~al.,}{{Wright}
  et~al.}{2010}]{wright_2010}
{Wright} E.~L.,  et~al., 2010, \mn@doi [\aj] {10.1088/0004-6256/140/6/1868},
  \href {https://ui.adsabs.harvard.edu/abs/2010AJ....140.1868W} {140, 1868}

\bibitem[\protect\citeauthoryear{{Zucker}, {Schlafly}, {Speagle}, {Green},
  {Portillo}, {Finkbeiner}  \& {Goodman}}{{Zucker} et~al.}{2018}]{zucker_2018}
{Zucker} C.,  {Schlafly} E.~F.,  {Speagle} J.~S.,  {Green} G.~M.,  {Portillo}
  S. K.~N.,  {Finkbeiner} D.~P.,   {Goodman} A.~A.,  2018, \mn@doi [\apj]
  {10.3847/1538-4357/aae97c}, \href
  {https://ui.adsabs.harvard.edu/abs/2018ApJ...869...83Z} {869, 83}

\bibitem[\protect\citeauthoryear{van~der Walt, Colbert  \& Varoquaux}{van~der
  Walt et~al.}{2011}]{vanderwalt11}
van~der Walt S.,  Colbert S.~C.,   Varoquaux G.,  2011, \mn@doi [Computing in
  Science and Engineering] {10.1109/MCSE.2011.37}, 13, 22

\makeatother
\end{thebibliography}
\input{ms.bbl}

%\begin{thebibliography}{99}
%\end{thebibliography}

%%%%%%%%%%%%%%%%%%%%%%%%%%%%%%%%%%%%%%%%%%%%%%%%%%

%%%%%%%%%%%%%%%%% APPENDICES %%%%%%%%%%%%%%%%%%%%%

\appendix
\section{RUWE Analysis}

\begin{figure*}
	\includegraphics[width=\textwidth]{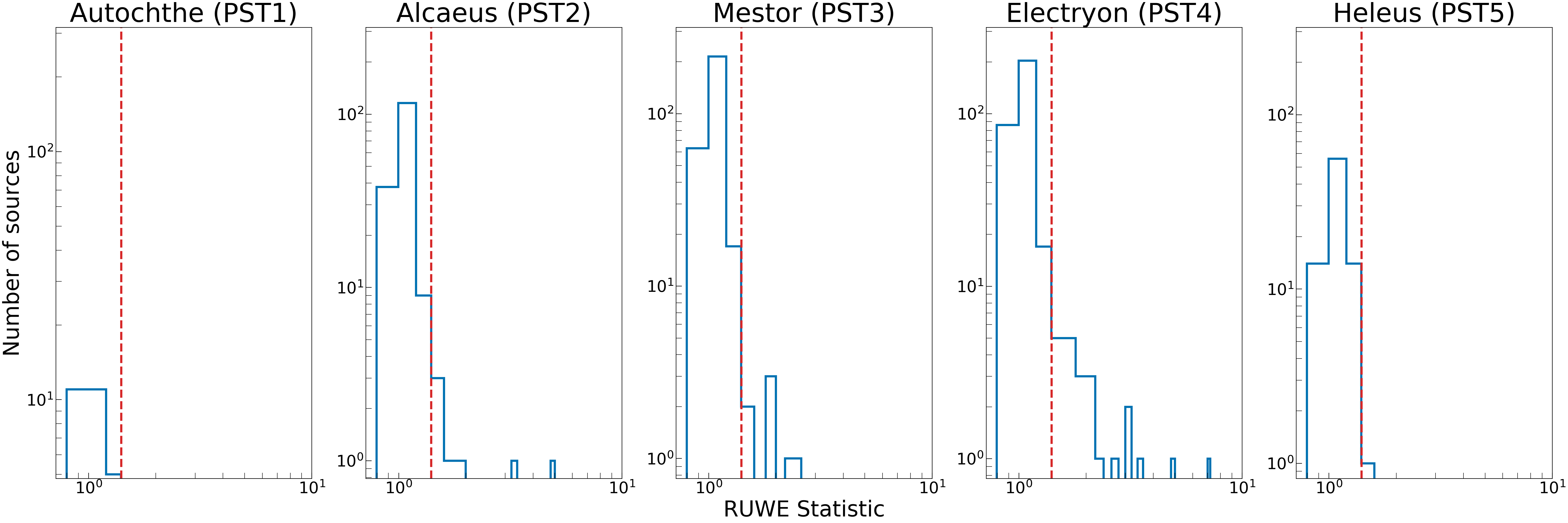}
    \caption{The RUWE histograms for the samples of young stars identified in this paper in log scale. Sources to the right of the red dashed line (RUWE of 1.4) might have problematic astrometric fits and could be astrometric binaries. For all groups, the fraction of stars above that threshold is low.}
    \label{fig:ruwe_hist}
\end{figure*}

In this Appendix we look into the RUWE (re-normalised unit weight error) statistic for the members of all five groups. Based on \citep{lindegren_2018}, a value of RUWE greater than 1.4 might indicate a non-single source or a source with an otherwise problematic astrometric fit. In Fig.~\ref{fig:ruwe_hist} we show the RUWE histograms for the five new groups of young stars identified in this paper. Although our samples do contain some sources with RUWE $>1.4$, the number of such sources is insignificant relative to the total number of stars in the samples (see Table \ref{tab:ruwe_stat}). The fraction of possible binaries with a high RUWE value is well below 5\% for all groups, except for Electryon (PST4), for which is still only 7\%. We note that high RUWE does not necessarily imply that the object is a binary. We conclude binaries do not have a significant impact on the findings in this paper.

\iffalse
\begin{table*}
\centering
\caption{RUWE > 1.4  Statistics}
\label{tab:ruwe_stat}
\begin{tabular}{l c | c c} % four columns, alignment for each
\hline
\hline
& $N_{total}$ & Number of Sources & Percentage (\%)\\
\hline

Autochthe (PST1) & 27 & 0 & 0 \\
Alcaeus (PST2) & 170 & 7 & 4.1 \\
Mestor (PST3) & 302 & 10 & 3.1 \\
Electryon (PST4) & 329 & 24 & 7.0\\
Heleus (PST5) & 85 & 1 & 1.1 \\

\hline
\hline

\end{tabular}
\end{table*}
\fi

%\iffalse
\begin{table*}
\centering
\caption{RUWE Statistics}
\label{tab:ruwe_stat}
\begin{tabular}{l c c c c c} % four columns, alignment for each
\hline
\hline
                                           & Autochthe (PST1) & Alcaeus (PST2) & Mestor (PST3) & Electryon (PST4) & Heleus (PST5)\\
\hline
Final sample size                   & 27            & 170               & 302           & 329               & 85   \\
Sources with RUWE > 1.4             & 0             & 7                 & 7             & 23                & 1  \\
Percentage with RUWE > 1.4 (\%)     & 0             & 4.1               & 2.3           & 7.0               & 1.2 \\

\hline
\hline

\end{tabular}
\end{table*}
%\fi

% ================================================================

\section{Lists of Members}
\label{appendixB}
In this Appendix we list in five separate tables, Tables \ref{tab:autochthe_members}-\ref{tab:heleus_members}, the members of each group along with their main properties, as they are listed in {\it Gaia} DR2, and their ID from \textit{AllWISE}. Rows with ``-" indicate that no WISE counterpart was found for the given source. At this stage these objects should be considered candidates.

\onecolumn

\begin{table*}
    \centering
	\caption{List of Autochthe (PST1) members}
    \label{tab:autochthe_members}
\begin{tabular}{ccccccccc} 
    \hline
    \hline

ID & \textit{Gaia} DR2 Designation & $\alpha$ / deg  & $\delta$ / deg & $\mu_\alpha^*$ / mas\,yr$^{-1}$ & $\mu_\delta$ / mas\,yr$^{-1}$  & $\varpi$ / mas & G / mag & ALLWISE Designation  \\  
    \hline
    
1 & 123910081367783552 & 50.7867 & 30.8073 & 7.78$\pm 0.36$ & -7.96$\pm 0.26$ & 3.58$\pm 0.27$ & 18.04 & J032308.82+304826.2 \\ 
2 & 123967049813899008 & 50.8526 & 31.0514 & 8.54$\pm 0.28$ & -7.89$\pm 0.17$ & 3.29$\pm 0.17$ & 17.02 & J032324.62+310305.2 \\ 
3 & 123990242637876480 & 51.3240 & 30.7741 & 7.62$\pm 0.37$ & -9.14$\pm 0.27$ & 3.49$\pm 0.25$ & 17.21 & J032517.75+304626.6 \\ 
4 & 123990517515783680 & 51.3938 & 30.7911 & 7.14$\pm 0.51$ & -8.68$\pm 0.32$ & 2.93$\pm 0.28$ & 18.11 & J032534.52+304727.9 \\ 
5 & 123990934127232768 & 51.2892 & 30.7724 & 7.94$\pm 0.59$ & -9.03$\pm 0.42$ & 3.67$\pm 0.39$ & 17.89 & - \\ 
6 & 123991277725418752 & 51.2941 & 30.8096 & 7.74$\pm 0.30$ & -9.30$\pm 0.22$ & 3.71$\pm 0.20$ & 16.79 & J032510.59+304834.6 \\ 
7 & 123993373669456000 & 51.3352 & 30.8657 & 8.63$\pm 0.93$ & -8.76$\pm 0.52$ & 3.13$\pm 0.44$ & 18.41 & J032520.44+305156.7 \\
\hline
    \hline
    \end{tabular}
\end{table*}

\iffalse
\begin{table*}
    \centering
	\caption{List of Autochthe (PST1) members}
    \label{tab:autochthe_members}
    \begin{tabular}{ccc cc ccc} 
    \hline
    \hline

ID & $\alpha$ / deg  & $\delta$ / deg & $\mu_\alpha^*$ / mas\,yr$^{-1}$ & $\mu_\delta$ / mas\,yr$^{-1}$  & $\varpi$ / mas & G / mag & ALLWISE Designation  \\  
    \hline
    
\input{table_auto.tex}

	\hline
    \hline
    \end{tabular}
\end{table*}
\fi

\clearpage
% ============================================
% Alcaeus Table
% ============================================
\renewcommand{\thefootnote}{\fnsymbol{footnote}}
\begin{small}
\begin{longtable}{ccccccccc}
\caption{List of Alcaeus (PST2) members.}\\
\hline

ID & \textit{Gaia} DR2 Designation & $\alpha$ / deg  & $\delta$ / deg & $\mu_\alpha^*$ / mas\,yr$^{-1}$ & $\mu_\delta$ / mas\,yr$^{-1}$  & $\varpi$ / mas & G / mag & ALLWISE Designation  \\  

\hline
\hline
\endfirsthead
\caption{continued.}\\

ID & \textit{Gaia} DR2 Designation & $\alpha$ / deg  & $\delta$ / deg & $\mu_\alpha^*$ / mas\,yr$^{-1}$ & $\mu_\delta$ / mas\,yr$^{-1}$  & $\varpi$ / mas & G / mag & ALLWISE Designation  \\  

\hline\hline
\endhead
\hline
\endfoot
\hline
\hline
1 & 168239783157120768 & 57.7202 & 30.6912 & 5.61$\pm 0.09$ & -9.33$\pm 0.06$ & 3.48$\pm 0.06$ & 9.57 & J035052.83+304128.2 \\ 
2 & 168240126754499200 & 57.7417 & 30.7094 & 6.82$\pm 0.08$ & -10.48$\pm 0.06$ & 2.11$\pm 0.05$ & 11.76 & J035058.00+304233.8 \\ 
3 & 168319875707029760 & 57.6629 & 31.2310 & 6.39$\pm 0.37$ & -9.33$\pm 0.19$ & 3.46$\pm 0.18$ & 17.61 & J035039.10+311351.6 \\ 
4 & 168320940858918656 & 57.6124 & 31.2577 & 6.41$\pm 0.48$ & -9.18$\pm 0.26$ & 3.26$\pm 0.23$ & 17.95 & - \\ 
5 & 168349528161427328 & 59.0219 & 30.3969 & 6.64$\pm 0.30$ & -9.01$\pm 0.16$ & 3.62$\pm 0.17$ & 17.58 & J035605.25+302348.8 \\ 
6 & 168365402358374400 & 59.3639 & 30.6261 & 5.22$\pm 0.60$ & -9.11$\pm 0.40$ & 4.05$\pm 0.35$ & 18.60 & J035727.30+303733.7 \\ 
7 & 168371655832862464 & 59.4851 & 30.6842 & 6.28$\pm 0.15$ & -9.06$\pm 0.10$ & 3.43$\pm 0.07$ & 16.16 & J035756.43+304103.0 \\ 
8 & 168394088445360000 & 58.7208 & 30.4448 & 6.12$\pm 0.65$ & -9.10$\pm 0.40$ & 2.54$\pm 0.33$ & 18.76 & J035452.98+302641.0 \\
\end{longtable}
\end{small}
% ============================================

\clearpage

% ============================================
% Mestor Table
% ============================================

\renewcommand{\thefootnote}{\fnsymbol{footnote}}
\begin{small}
\begin{longtable}{ccccccccc}
\caption{\label{tab:mestor_members} List of Mestor (PST3) members.}\\
\hline

ID & \textit{Gaia} DR2 Designation & $\alpha$ / deg  & $\delta$ / deg & $\mu_\alpha^*$ / mas\,yr$^{-1}$ & $\mu_\delta$ / mas\,yr$^{-1}$  & $\varpi$ / mas & G / mag & ALLWISE Designation  \\  

\hline
\hline
\endfirsthead
\caption{continued.}\\

ID & \textit{Gaia} DR2 Designation & $\alpha$ / deg  & $\delta$ / deg & $\mu_\alpha^*$ / mas\,yr$^{-1}$ & $\mu_\delta$ / mas\,yr$^{-1}$  & $\varpi$ / mas & G / mag & ALLWISE Designation  \\   

\hline\hline
\endhead
\hline
\endfoot
\hline
\hline
1 & 217206464781552896 & 58.1075 & 33.4541 & 3.37$\pm 0.09$ & -4.70$\pm 0.05$ & 2.48$\pm 0.04$ & 14.95 & J035225.80+332714.9 \\ 
2 & 217207667372389888 & 58.0768 & 33.5165 & 2.61$\pm 0.30$ & -5.54$\pm 0.18$ & 2.95$\pm 0.15$ & 17.62 & J035218.43+333059.3 \\ 
3 & 217211687461582464 & 57.8683 & 33.4296 & 3.44$\pm 0.24$ & -4.74$\pm 0.14$ & 2.22$\pm 0.11$ & 16.26 & J035128.39+332546.6 \\ 
4 & 217215329593853056 & 57.7544 & 33.4747 & 3.19$\pm 0.41$ & -4.99$\pm 0.26$ & 2.35$\pm 0.19$ & 18.16 & J035101.05+332829.1 \\ 
5 & 217215535752283520 & 57.7521 & 33.4752 & 3.17$\pm 0.46$ & -5.00$\pm 0.28$ & 2.41$\pm 0.21$ & 18.21 & J035100.51+332830.7 \\ 
6 & 217223026175137152 & 58.0600 & 33.6819 & 3.50$\pm 0.17$ & -4.76$\pm 0.11$ & 2.55$\pm 0.09$ & 15.84 & J035214.41+334054.8 \\ 
7 & 217241614793772800 & 57.3857 & 33.4518 & 3.00$\pm 0.29$ & -4.55$\pm 0.18$ & 2.43$\pm 0.14$ & 17.60 & J034932.55+332706.6 \\ 
8 & 217251166797795968 & 57.0772 & 33.5171 & 3.98$\pm 0.65$ & -4.66$\pm 0.49$ & 2.41$\pm 0.36$ & 19.05 & J034818.53+333101.2 \\
\end{longtable}
\end{small}
% ============================================

\clearpage

% ============================================
% Electryon Table
% ============================================

\renewcommand{\thefootnote}{\fnsymbol{footnote}}
\begin{small}
\begin{longtable}{ccccccccc}
\caption{\label{tab:electryon_members} List of Electryon (PST4) members.}\\
\hline

ID & \textit{Gaia} DR2 Designation & $\alpha$ / deg  & $\delta$ / deg & $\mu_\alpha^*$ / mas\,yr$^{-1}$ & $\mu_\delta$ / mas\,yr$^{-1}$  & $\varpi$ / mas & G / mag & ALLWISE Designation  \\   

\hline
\hline
\endfirsthead
\caption{continued.}\\

ID & \textit{Gaia} DR2 Designation & $\alpha$ / deg  & $\delta$ / deg & $\mu_\alpha^*$ / mas\,yr$^{-1}$ & $\mu_\delta$ / mas\,yr$^{-1}$  & $\varpi$ / mas & G / mag & ALLWISE Designation  \\

\hline\hline
\endhead
\hline
\endfoot
\hline
\hline
1 & 167597805805592832 & 59.7788 & 30.3499 & 3.57$\pm 0.30$ & -6.19$\pm 0.18$ & 2.85$\pm 0.14$ & 17.14 & J035906.91+302059.7 \\ 
2 & 167635120482381824 & 60.5850 & 30.4189 & 3.99$\pm 0.31$ & -6.25$\pm 0.16$ & 2.81$\pm 0.15$ & 17.31 & J040220.40+302508.2 \\ 
3 & 167640888619905152 & 60.2569 & 30.3142 & 4.32$\pm 0.34$ & -6.40$\pm 0.21$ & 2.84$\pm 0.17$ & 17.93 & J040101.66+301851.0 \\ 
4 & 167644775566093952 & 60.4771 & 30.3674 & 3.17$\pm 1.28$ & -6.60$\pm 0.70$ & 2.84$\pm 0.67$ & 20.01 & J040154.50+302202.8 \\ 
5 & 167650543706474112 & 60.3216 & 30.5139 & 3.70$\pm 0.30$ & -6.36$\pm 0.17$ & 2.90$\pm 0.14$ & 17.73 & J040117.18+303050.0 \\ 
6 & 167652197272316544 & 60.5110 & 30.6033 & 3.08$\pm 0.36$ & -6.24$\pm 0.21$ & 2.75$\pm 0.19$ & 17.90 & J040202.65+303611.9 \\ 
7 & 167653434222887552 & 60.4767 & 30.6627 & 4.12$\pm 0.30$ & -5.58$\pm 0.16$ & 2.85$\pm 0.14$ & 17.66 & J040154.40+303945.6 \\ 
8 & 167656560959245184 & 60.9303 & 30.5051 & 3.54$\pm 0.41$ & -5.60$\pm 0.20$ & 3.01$\pm 0.18$ & 17.73 & J040343.25+303018.3 \\
\end{longtable}
\end{small}
% ============================================

\clearpage

% ============================================
% Heleus Table
% ============================================

\renewcommand{\thefootnote}{\fnsymbol{footnote}}
\begin{small}
\begin{longtable}{ccccccccc}
\caption{\label{tab:heleus_members} List of Heleus (PST5) members.}\\
\hline

ID & \textit{Gaia} DR2 Designation & $\alpha$ / deg  & $\delta$ / deg & $\mu_\alpha^*$ / mas\,yr$^{-1}$ & $\mu_\delta$ / mas\,yr$^{-1}$  & $\varpi$ / mas & G / mag & ALLWISE Designation  \\  

\hline
\hline
\endfirsthead
\caption{continued.}\\

ID & \textit{Gaia} DR2 Designation & $\alpha$ / deg  & $\delta$ / deg & $\mu_\alpha^*$ / mas\,yr$^{-1}$ & $\mu_\delta$ / mas\,yr$^{-1}$  & $\varpi$ / mas & G / mag & ALLWISE Designation  \\  

\hline\hline
\endhead
\hline
\endfoot
\hline
\hline
1 & 71961184908996480 & 56.7505 & 29.1889 & 3.40$\pm 0.30$ & -5.60$\pm 0.18$ & 2.29$\pm 0.17$ & 17.01 & J034700.13+291119.9 \\ 
2 & 71986576755460224 & 56.5719 & 29.5946 & 2.45$\pm 0.08$ & -5.27$\pm 0.04$ & 2.29$\pm 0.04$ & 14.68 & J034617.26+293540.5 \\ 
3 & 72003790982983168 & 56.0256 & 29.3768 & 3.80$\pm 0.90$ & -5.72$\pm 0.62$ & 2.03$\pm 0.54$ & 19.31 & J034406.13+292236.5 \\ 
4 & 72026227893573248 & 56.2939 & 29.5134 & 2.48$\pm 0.25$ & -5.12$\pm 0.12$ & 2.36$\pm 0.14$ & 16.21 & J034510.54+293048.0 \\ 
5 & 72030071888336128 & 56.2984 & 29.6047 & 3.07$\pm 0.63$ & -5.93$\pm 0.33$ & 2.59$\pm 0.33$ & 18.06 & J034511.61+293616.7 \\ 
6 & 72032859323056640 & 56.4262 & 29.5752 & 3.02$\pm 0.54$ & -5.68$\pm 0.26$ & 2.46$\pm 0.40$ & 18.12 & J034542.28+293430.6 \\ 
7 & 72033271639908864 & 56.4305 & 29.6296 & 2.58$\pm 0.32$ & -5.23$\pm 0.17$ & 2.25$\pm 0.15$ & 17.04 & J034543.31+293746.4 \\ 
8 & 72035092706027520 & 56.5297 & 29.6830 & 2.78$\pm 0.15$ & -5.65$\pm 0.10$ & 2.48$\pm 0.09$ & 15.54 & - \\
\end{longtable}
\end{small}
% ============================================

%%%%%%%%%%%%%%%%%%%%%%%%%%%%%%%%%%%%%%%%%%%%%%%%%%

% Don't change these lines
\bsp	% typesetting comment
\label{lastpage}
\end{document}